\documentclass[a4paper,onecolumn,10pt,accepted=2024-11-24]{quantumarticle}
\pdfoutput=1
\usepackage[utf8]{inputenc}
\usepackage[english]{babel}
\usepackage[T1]{fontenc}
\usepackage[usenames,dvipsnames]{xcolor}
\usepackage{amssymb}
\usepackage{graphicx}
\usepackage{amsmath}
\usepackage{mathtools}  
\mathtoolsset{showonlyrefs} 
\usepackage{dsfont}
\usepackage[bookmarks=true,colorlinks,citecolor=blue,urlcolor=blue]{hyperref}
\usepackage{physics}
\usepackage[numbers]{natbib}

\usepackage{bm}

\usepackage[utf8]{inputenc}
\usepackage[linesnumbered,ruled,vlined]{algorithm2e}
\SetKwInput{KwInput}{Input}                
\SetKwInput{KwOutput}{Output}

\begin{document}

\author{Elisabeth Wybo}
\email{elisabeth.wybo@meetiqm.com}
\affiliation{IQM Quantum Computers, Georg-Brauchle-Ring 23-25, 80992 München, Germany}
\author{Martin Leib}
\email{martin.leib@meetiqm.com}
\affiliation{IQM Quantum Computers, Georg-Brauchle-Ring 23-25, 80992 München, Germany}

\title{Vanishing performance of the parity-encoded quantum approximate optimization algorithm applied to spin-glass models}

\begin{abstract}
    The parity mapping provides a geometrically local encoding of the Quantum Approximate Optimization Algorithm (QAOA), at the expense of having a quadratic qubit overhead for all-to-all connected problems. In this work, we benchmark the parity-encoded QAOA on spin-glass models. We address open questions in the scaling of this algorithm. In particular, we show that for fixed number of parity-encoded QAOA layers, the performance or the output energy, vanishes towards zero (the value achieved
by random guessing) with problem size $N$ as $N^{-1/2}$. Our results suggest that the parity-encoded QAOA does not have a promising scaling compared to the standard version of QAOA. We perform tensor-network calculations to confirm our results, and comment on the concentration of optimal QAOA parameters over problem instances. 
\end{abstract}

\maketitle

\section{Introduction}

Variational quantum algorithms are considered valuable candidates for providing useful quantum-computational results on near-term devices~\cite{Cerezo2020,Bharti2021}. The Quantum Approximate Optimization Algorithm (QAOA), originally introduced by Farhi in Ref.~\cite{Farhi2014}, is such a variational algorithm for gate-based machines, that has as goal to solve combinatorial optimization problems. In some specific cases, QAOA has provable performance guarantees, and can even outperform the best known assumption-free classical algorithm, without the need of optimizing the variational parameters on an instance-by-instance basis~\cite{Farhi2022,Basso2021}. 

Most combinatorial optimization problems, however, are defined on arbitrary graph topologies, some even all-to-all connected. This implies that the topology of the problem rarely matches the topology of the actual device, and that rerouting using \texttt{SWAP} networks is necessary~\cite{OGorman2019,Lotshaw2022}. A general way to circumvent the need for rerouting, is to employ different problem encodings that are native to the topology of the hardware, for instance a square grid. In this respect, the parity encoding~\cite{Lechner2015} forms a general encoding scheme to map geometrically non-local optimization problems to local ones. The encoded `parity qubits' represent products of logical qubits, at the expense of having a quadratic overhead and a set of local constraints that need to be satisfied. 

However, apart from this clear advantage that qubit routing is not needed, it is unclear if such a strategy is actually promising in terms of algorithmic performance. In this work, we aim to address this question by investigating the scaling of the performance of parity-encoded QAOA as introduced in Ref.~\cite{Lechner2018} for paradigmatic all-to-all connected problems, namely Sherrington-Kirkpatrick spin glasses. 
Our work addresses for the first time the open questions about the scaling of this algorithm, and also connects to the question of whether non-local optimization problems can be solved up to high precision by geometrically local algorithms. Thus also addressing of whether such algorithms can be useful for near-term hardware implementations. While formally not excluding a faster runtime, our results suggests that there are no clear scaling advantages of the parity-encoded algorithm compared to the standard QAOA algorithm, that would justify the quadratic qubit overhead. As we will demonstrate, the main reason for this, is that the number of parity QAOA layers $p$ needs to grow with the problem size in order to have non-vanishing performance in the thermodynamic limit. Here, we define performance as the average energy of decoded bitstrings according to `qubit lines', in close correspondence to the decoding scheme proposed in Ref.~\cite{Weidinger2023}. No matter how the performance is exactly defined, it is likely that $p$ will need to grow with growing system size, as each parity-encoded QAOA layer has fixed depth because of the geometric locality of the algorithm. This should be contrasted with the standard (`vanilla') QAOA where the depth of each QAOA layer scales with system size for all-to-all connected problems. 

Additionally, we perform numerical experiments based on tensor-network techniques. These techniques have become standard to simulate quantum algorithms~\cite{Shi2008,Huang2021,Gray2020,Liu2022,Gray2022}. These experiments confirm our findings and illustrate that the algorithm does not give promising results. In practice, we even find that it is not outperforming a classical algorithm that simply sets each parity spin according the sign of the interaction strength. Importantly, however, QAOA is considered to be a hybrid quantum-classical algorithm~\cite{McClean2015,Endo2020} because the variational parameters in the ansatz circuit need to be optimized classically. This is known to be a tedious procedure, that is affected by various difficulties, e.g. Barren plateaus~\cite{Cerezo2020a}. However, it has become clear that a parameter concentration phenomenon is present for vanilla QAOA, e.g. for MaxCut or Sherrington-Kirkpatrick spin-glass  problems~\cite{Farhi2022,Basso2021,Brandao2018,Streif2020,Akshay2021,Galda2021,Sack2021,Galda2023}. This means that the optimal variational parameters for typical problem instances concentrate around specific values. If the optimal parameters do not depend much on system size (which is the case for Sherrington-Kirkpatrick vanilla QAOA), such a parameter concentration is of great practical importance, as expensive parameter optimization loops can be eliminated, even for large instances. Our numerical findings indicate that also for the parity-encoded QAOA the optimal parameters concentrate. For $p=1$, we determine these optimal angles analytically. However, as our scaling results indicate, unlike vanilla QAOA for Sherrington-Kirkpatrick~\cite{Farhi2022}, fixed optimal angles do not lead to fixed performance for parity-encoded QAOA. This follows from the fact that parity-constraints `average out' in the optimization procedure, preventing entanglement to build up. In the end, we will conclude that the parity-encoded QAOA does not show promising performance when compared to vanilla QAOA. 

The structure of this paper is the following: in Sec.~\ref{sec:setup} we introduce the problem we want to solve and give an overview of parity-encoded QAOA. For determining the quality of the solution, we use a similar cost function as in Ref.~\cite{Weidinger2023}. Second, in Sec.~\ref{sec:an} we discuss our analytical results, and give a vanishing bound on the performance of parity-encoded QAOA for fixed number of layers. In Sec.~\ref{sec:tn}, we confirm our findings by tensor-network calculations of the circuit, and finally conclude in Sec.~\ref{sec:concl}.

\section{Parity encoded QAOA and spin-glass models} \label{sec:setup}

\begin{figure}
\centering
\includegraphics[width=0.6\textwidth]{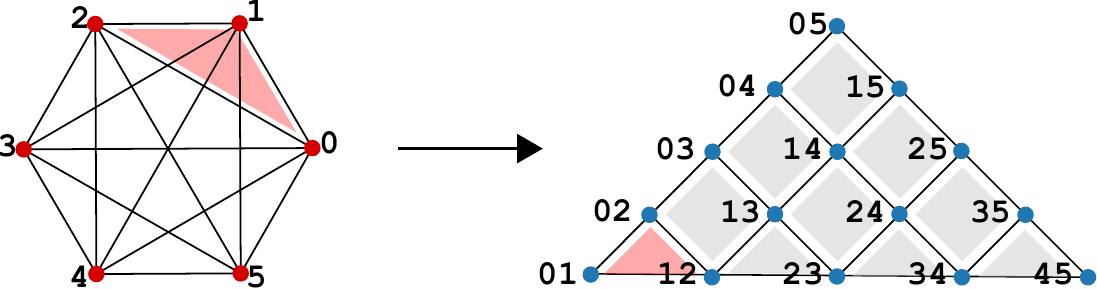}
\caption{Sketch of the parity mapping: the parity qubits (shown in blue) represent the parity between two logical qubits (shown in red) of the original problem graph, consisting of $N=6$ qubits. A problem graph that is all-to-all connected, leads to a triangular layout on a square lattice of $K=N(N-1)/2$ parity qubits, where the parity qubits need to satisfy plaquette constraints. Those constraints follow from a set of closed cycles that form a cycle basis of the original problem graph. An example of such a cycle is indicated in red.  }
\label{fig:lhz_mapping}
\end{figure}

We consider a paradigmatic classical all-to-all connected spin-glass model, known as the Sherrington-Kirkpatrick model~\cite{Sherrington1975}
\begin{equation} \label{eq:SK_log}
    H_P = \sum_{i=0}^{N-2} \sum_{j=i+1}^{N-1}  J_{ij} Z_i Z_j  
\end{equation}
defined on $N$ spin variables. For simplicity we will usually take $J_{ij} \in \{ -1,1\}$, but our scaling results just rely on the fact that the couplings are taken from a symmetric distribution. This model is important in physics because it accurately describes glassy phases of matter. It is however also a model of strong mathematical interest because some exact results can be derived~\cite{Talagrand2011,Talagrand2011_2}. However, finding the ground state of large spin-glass instances belongs to the class of NP-hard problems, meaning that there is no quantum or classical algorithm expected that can find the exact ground state in polynomial time.  Recently however, under the assumption that this model has full replica-symmetry breaking, a very efficient approximate message-passing algorithm has been developed in Ref.~\cite{Montanari2018}. This algorithm finds a solution that is $1-\epsilon$ away from the optimum with high probability. On the other hand, conventional classical semi-definite programming methods have provable, and assumption-free, average performance guarantees~\cite{Aizenman1987,Montanari2015,Bandeira2019}. Additionally, Ref.~\cite{MisraSpieldenner2023} proposed a `quantum-inspired' classical algorithm based on a mean-field version of QAOA, and found good average performance. On the quantum side Refs.~\cite{Farhi2022,Basso2021} show that vanilla QAOA can outperform these assumption-free conventional methods at $p=11$. When complemented with a `relax-and-round' approach, this even happens for smaller-depth QAOA, see Ref.~\cite{Dupont2023}.

On the other hand, the approximate message-passing algorithm discussed in Ref.~\cite{Montanari2018} with complexity $f(\epsilon)N^2$, leaves little to no room for a quantum computational advantage in the sense of improving the solution quality with a polynomial-time quantum algorithm. However, an interesting open question is whether a quantum polynomial speedup can be shown convincingly. For this reason, the Sherrington-Kirkpatrick model remains an extremely interesting model to investigate with quantum approaches.

The potential quantum speedup combined with the vast amount of knowledge about this model from many different fields, make this model ideal for benchmarking purposes. In addition, it is particularly suitable for benchmarking the parity-encoded QAOA because of its all-to-all connectivity. In this case, the parity qubits have a fixed geometric structure, as we explain next.

In the parity encoding, each parity qubit represents the product of two logical qubits $Z_i Z_j \leftrightarrow \sigma^z_{ij}$, hence an edge in the original problem graph. Thus the parity encoding of the problem consists out of $K= N (N-1)/2$ parity qubits that we will label by the combined index $ij$ with $i<j$ [see Fig.~\ref{fig:lhz_mapping}]. However, because of the redundancy of this encoding, the parity qubits need to satisfy a set of constraints. It can easily be seen that on closed cycles in the logical problem graph the parity operators $\sigma^z_{ij}$, expanded in their respective logical operators, multiply up to identity. In order to have a mapping between the low-energy states of the logical Hamiltonian~\eqref{eq:SK_log}, and the low-energy states of the encoded Hamiltonian, a subset of closed cycles needs to be added to the encoded Hamiltonian as penalty terms. The number of closed cycles in the problem graph is exponential $\mathcal{O}(N^N)$, but it is sufficient to pick a subset of cycles that forms a cycle basis. (Note that for an all-to-all connected $N$-node graph a cycle basis consist out of $P=(N-1)(N-2)/2$ elements.) A particular convenient choice of cycle basis can be represented by the triangular layout on a square lattice, where the square or triangluar plaquettes form the elements of the cycle basis. This is illustrated in Fig.~\ref{fig:lhz_mapping} for $N=6$ logical qubits~\footnote{Note that one could find a more compact cycle basis, the so-called fundamental cycle basis. A cycle basis is defined as fundamental if and only if every basis cycle contains an edge that is not included in any other cycle of the basis. However, these fundamental cycles can not be arranged in a convenient lattice structure.}. This structure is also referred to as LHZ triangle, after the authors of the original work in which this mapping was introduced, see Ref.~\cite{Lechner2015}.

Under the parity mapping, the problem Hamiltonian~\eqref{eq:SK_log} then takes the geometrically local Parity-Encoded (PE) form 
\begin{equation} \label{eq:H_PC}
    H_{PE} = \sum_{i=0}^{N-2} \sum_{j=i+1}^{N-1} J_{ij} \sigma^z_{ij} 
- \sum_{i=0}^{N-3} \sum_{j=i+1}^{N-2} C_{ij} \sigma^z_{ij} \sigma^z_{ij+1} \sigma^z_{i+1j+1}  [\sigma^z_{i+1j}]
\end{equation}
The first term encodes the problem in the coupling strengths $J_{ij}$, the second term are energy penalties for unsatisfied constraints. While these can in principle be fine-tuned for each plaquette, we will simply put $C_{ij}\equiv C>0$ in the remaining of this paper.

The structure of the QAOA circuit corresponding to this Hamiltonian is discussed in Ref.~\cite{Lechner2018}. In particular, each layer of the QAOA contains three unitaries, as there are the constraints in addition to the conventional `problem' and `mixer' layers. The parity QAOA ansatz is given by
\begin{equation}\label{eq:psi}
 \ket{\psi(\bm{\gamma},\bm{\beta}, \bm{\Omega})} =  U_x(\beta_p) U_z(\gamma_p) U_c(\Omega_p)  \dots U_x(\beta_1) U_z(\gamma_1) U_c(\Omega_1) \ket{+}
\end{equation} 
  where $p$ is the number of layers and $\ket{+} = \bigotimes_{i=0}^{K-1}\frac{1}{\sqrt{2}}(\ket{0}+\ket{1}) $. The unitaries are defined as
  \begin{align}
  \begin{split} \label{eq:pqaoa_gates}
    &U_x(\beta) = \prod_{i<j} e^{-i \beta \sigma^x_{ij}}, \quad \\
    &U_z(\gamma) = \prod_{i<j} e^{-i \gamma J_{ij} \sigma^z_{ij}}, \quad \\
    &U_c(\Omega) = \prod_{i<j} e^{-i \Omega \sigma^z_{ij} \sigma^z_{ij+1} \sigma^z_{i+1j+1}  \sigma^z_{i+1j} }. 
    \end{split}
  \end{align}
We note here that both the mixer and problem layers are strictly local. The constraint layer contains geometrically local three-or four-body interactions, and generates entanglement. Conventionally, such a single four-body gate is decomposed as a depth-6 \texttt{CNOT}-ladder with an \texttt{Rz} rotation in the middle~\cite{Lechner2018}. However, for a full layer of these operators applied to qubits arranged on a square lattice, smaller depth decompositions are possible, see Ref.~\cite{Sriluckshmy2023}.   

The parametric angles need to be classically optimized in such a way that the energy
\begin{equation} \label{eq:E}
E(\bm{\gamma},\bm{\beta}, \bm{\Omega})  = \ev{H_{PE}}{\psi(\bm{\gamma},\bm{\beta}, \bm{\Omega})}
\end{equation}
is minimized. The angles satisfy certain symmetries that are discussed in Appendix~\ref{app:angle_sym}. However, this energy does not directly relate to the energy of a decoded bitstring, i.e. we can not directly associate a logical `solution' bitstring to a measured parity-encoded bitstring, as the parity-encoded bitstring will generally not satisfy all constraints. Indeed, we start from the $\ket{+}$ state which on average violates half of the constraints, and apply a QAOA circuit that in principle allows for the full exploration of the parity-encoded Hilbert space. 

 \begin{figure}
    \centering
    \includegraphics[width=0.6\textwidth]{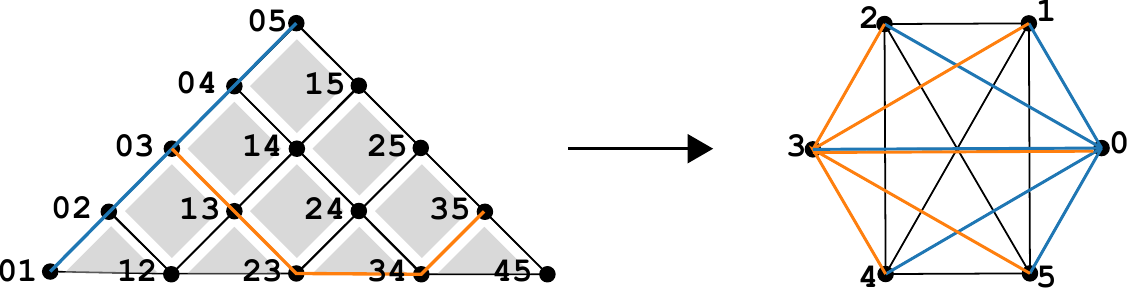}    \caption{A `qubit line' in the LHZ triangle is connecting parity qubits that each have one repeated index. Naturally, there are $N$ such qubit lines, and each of them corresponds to a spanning tree of fixed topology in the logical picture. Here, as an example, we show two such qubit lines with their corresponding spanning trees in blue and orange.}
    \label{fig:LHZ_ST}
\end{figure}

Applying a constraint-fixing procedure to explicitly correct broken constraints in the measured output of the algorithm, can potentially be computationally demanding. Furthermore, to understand the performance of the quantum algorithm, it is preferred not to add any expensive post-processing steps. Otherwise, they could dominate the scaling of the algorithm. 

Therefore, to measure the performance of the algorithm, we will evaluate the cost function~\cite{Weidinger2023}
\begin{equation} \label{eq:H_ST}
    H_{ST} = \frac{1}{N} \sum_{t=0}^{N-1} \sum_{i<j} J_{ij} \sigma^z_{it} \sigma^z_{tj},
\end{equation}
 on the output state. Here we use the notation $\sigma^z_{tj}\equiv \sigma^z_{jt}$, and $\sigma^z_{jj}=1$. This cost function evaluates the mean logical energy of decoding along the `qubit lines', which are the lines in the LHZ triangle with a common index. There are $N$ such lines, and they correspond to spanning trees in the logical problem, as sketched in Fig.~\ref{fig:LHZ_ST}. Thus if we pick such a spanning tree $t$ at random, we can associate a logical bitstring with a measured parity-encoded bitstring $\bm{z}$ in the following way: (i) We fix the logical qubit $t$, because of the $\mathbb{Z}_2$ symmetry of~\eqref{eq:SK_log} this is an arbitrary choice. (ii) We fix the remainder of logical qubits $i\neq t$ according to $\sigma^z_{it} = \sigma^z_{ti}$ measured in the parity bitstring. For example, if we choose the logical qubit $t$ to be in the `1' state, and we have the measurement outcome $\ev{\sigma^z_{it}}{\bm{z}}=-1$ in the parity bitstring, we then have that the $i$th qubit is in the `$-1$' state in the logical bitstring. In this way, we can associate $N$ logical bitstrings, each corresponding to one of the qubit line spanning trees, to the parity bitstring $\bm{z}$. If no constraints are broken in $\bm{z}$, these $N$ logical bitstrings are the same. 
 
 Following this procedure, Eq.~\eqref{eq:H_ST} gives the average logical energy of decoding along $N$ qubit lines. Broken constraints will lead to inconsistent logical bitstrings along different qubit lines. However, the exact ground state of Eq.~\eqref{eq:H_ST} encodes the logical solution, and decoding along any spanning tree will lead to this solution.
 
 The cost function~\eqref{eq:H_ST} originates from Ref.~\cite{Weidinger2023}, we refer to this work for further details. However, as we are interested in the average scaling of the algorithm, we will not explicitly sample bitstrings along qubit lines (or along more general spanning trees in the logical graph).

 In this paper, we define the performance ratio of parity-encoded QAOA as
 \begin{equation} \label{eq:P_ST}
     \mathcal{P}_{ST}(\bm{\gamma},\bm{\beta}, \bm{\Omega}) = -\frac{\ev{H_{ST}}{\psi(\bm{\gamma},\bm{\beta}, \bm{\Omega})}}{N^{3/2}}.
 \end{equation}
This quantity corresponds to the negative energy density of the logical problem. The renormalized logical Hamiltonian $\tilde{H}_P = \frac{1}{\sqrt{N}} H_P$ is extensive (i.e. its ground state energy scales linearly with $N$), thus further dividing it by $N$ gives the energy density. This explains the factor $N^{3/2}$ in the denominator. This quantity converges to the Parisi constant in the large $N$ limit  
 \begin{equation} \label{eq:parisi_limit}
     \lim_{N\rightarrow\infty} \mathcal{P}_{ST}^0 =  -\frac{\ev{H_{ST}}{\psi_0}}{N^{3/2}} = \epsilon_P = 0.763\dots,
 \end{equation}
 for the ground state $\ket{\psi_0}$ of typical instances. Note that this is then also trivially true for the average $\mathbb{E}_J(\mathcal{P}_{ST}^0)$ over all instances. Throughout this paper, we will mostly be interested in computing $\mathbb{E}_J(\mathcal{P}_{ST}(\bm{\gamma},\bm{\beta}, \bm{\Omega}))$.

\begin{figure*}
\centering
\includegraphics[width=0.6\textwidth]{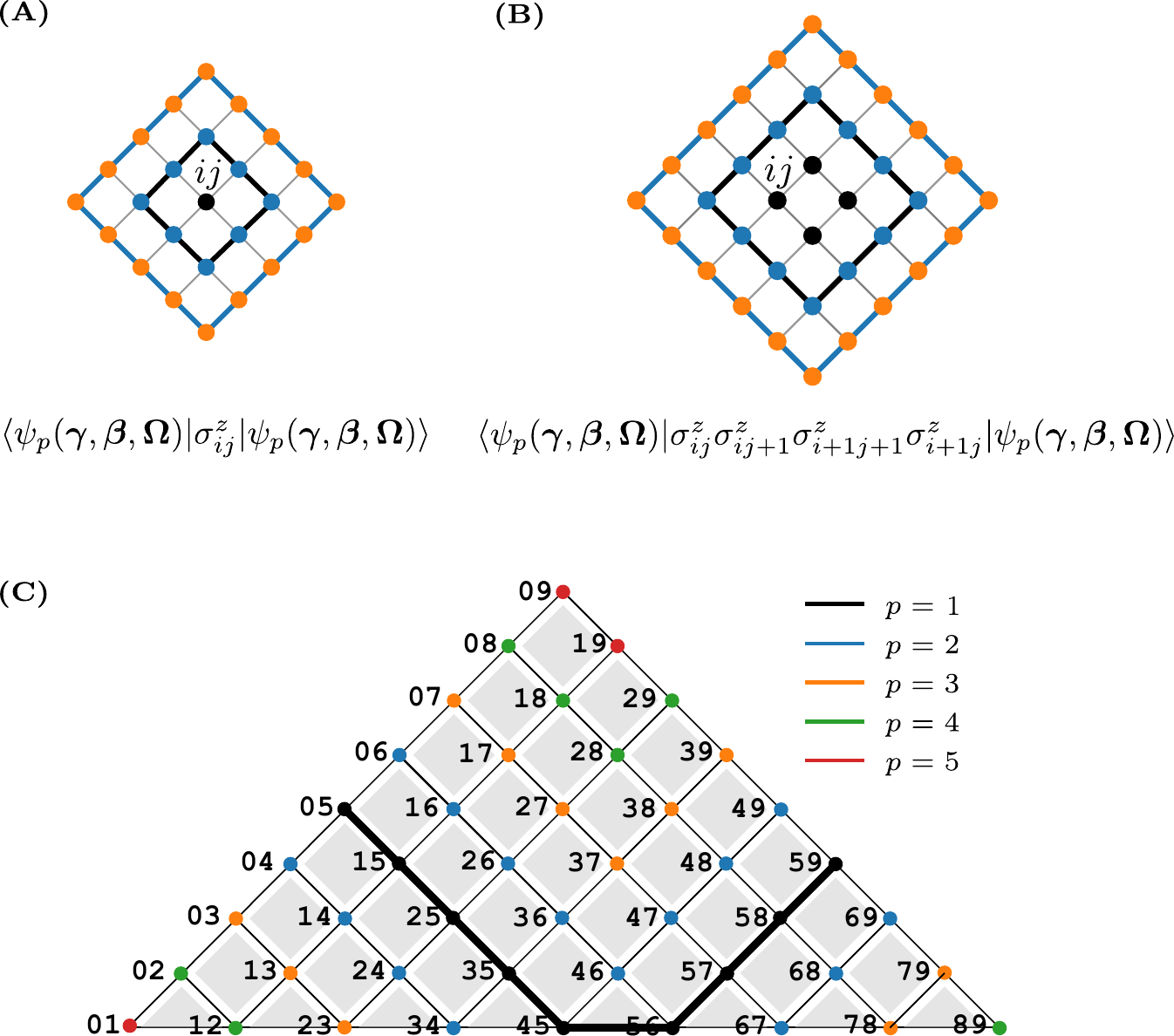}
\caption{(A) The bulk reverse causal cone of the onsite operator. The thick lines represent the boundary of the reverse causal cones, respectively shown for $p=1$ (black) and $p=2$ (blue). These reverse causal cones do not depend on the coupling strengths (the $J$'s) associated with the qubits on the boundary, but only on those excluding the boundary. We indicate this reduced dependence by coloring the qubits for which the coupling strength is relevant. E.g. for $p=1$ the $\sigma^z_{ij}$ RCC depends only on the coupling strength of the parity qubit indicated in black, for $p=2$ it depends on those indicated in black and blue, and so on. (B) The bulk reverse causal cone of the plaquette operator. (C) Sketch of the LHZ triangle for $N=10$ with the $t=5$ qubit line indicated in black. For this qubit line, the sites $ij$ for which the RCC of $\sigma^z_{i5} \sigma^z_{5j}$ depends on $J_{ij}$ are indicated by the color scheme as a function of $p$.}
\label{fig:RCCs}
\end{figure*}

\section{Bounding the performance of parity-encoded QAOA} \label{sec:an}
In this section, we discuss our analytic results in evaluating the performance of parity-encoded QAOA. In the spirit of Ref.~\cite{Farhi2022}, we will evaluate the value of the cost function averaged over all problem instances. Due to the lengthy calculations involved, we have only done this explicitly for $p=1$. However, we can derive asymptotically tight upper bounds on the performance~\eqref{eq:P_ST} of parity-encoded QAOA for arbitrary $p$. 

Because of the geometric locality of the parity-encoded algorithm [see Eq.~\eqref{eq:pqaoa_gates}], we can perform the calculations on the light cones or `Reverse Causal Cones' (RCCs) of the observables in the Hamiltonian [see Fig.~\ref{fig:RCCs}]. In case $p$ is small enough compared to $N$, i.e. when $p<N-1$, these RCCs do not span the whole system. This RCC reduction is due to unitary cancellations of gates that commute with the observables. The RCC concept was already included in the original QAOA paper~\cite{Farhi2014}, but is for standard QAOA applied on spin-glass models not very useful~\cite{Farhi2022}, as each RCC spans the whole system because of the all-to-all connectivity.

\subsection{Results for $p=1$}

For $p=1$ parity-encoded QAOA we can derive the following result 
\begin{multline} \label{eq:ex_HPE}
        \mathbb{E}_J \left( \ev{ H_{PE}}{\psi(\gamma,\beta, \Omega)} \right) \approx 
     K\cos^4(2\Omega) \sin(2\beta) \sin(2\gamma) \\
     - 4\ P C \cos(2\gamma) \sin(2\Omega) \cos^3(2\Omega) \sin(2\beta) \cos(2\beta) 
     \left( \cos^2(2\beta) - \cos^2(2\gamma) \cos^2(2\Omega) \sin^2(2\beta) \right),
\end{multline}
where $P=(N-1)(N-2)/2$ is the number of constraints, or plaquette terms in Eq.~\eqref{eq:H_PC}. The full derivation is included in Appendix~\ref{app:calc_p1}.
Here, the approximation assumes that all reverse causal cones have the same square shape as shown in Fig.~\ref{fig:RCCs}{(A,B)}. This is true for the bulk of the reverse causal cones $\mathcal{O}(N^2)$. Boundary effects $\mathcal{O}(N)$ because of different shapes can be neglected. Concise expressions for $p=1$ vanilla QAOA have also been derived for various problems including spin glasses~\cite{Ozaeta2020}.

\begin{figure}
    \centering
    \includegraphics[width=0.6\textwidth]{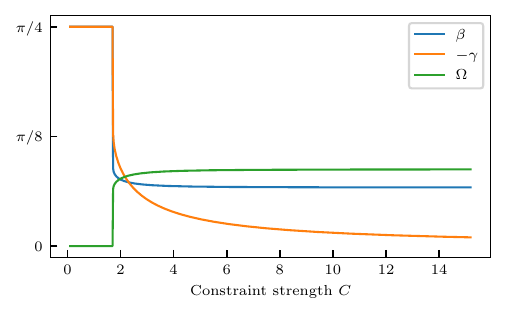}
    \caption{\textbf{Optimal parity-encoded QAOA angles} for $p=1$. There is a transition from trivial to non-trivial angles as the constraint strength becomes larger than the gap. Here the couplings are $J_{ij} \in \{-1,1\}$. In the large $C$ limit, the angle that drives the problem Hamiltonian vanishes.}
    \label{fig:angles_C}
\end{figure}

Eq.~\eqref{eq:ex_HPE} can be minimized as a function of the angles. In Fig.~\ref{fig:angles_C}, we show the resulting optimal angles as a function of the constraint strength $C$, for large $N$. As expected, the constraint strength should be taken larger than the gap to obtain non-trivial angles. For $J_{ij} \in \{-1,1\}$, the gap is 2 for non-degenerate problems. This corresponds to the energy difference induced by flipping spins on an edge in the ground state configuration. We also observe that if the constraint strength is taken very large, the problem angle becomes trivial. This indicates that the algorithm gradually forgets about the specific problem it is supposed to solve, and only tries to satisfy the constraints.

Next, we want to calculate the average optimal performance $\mathbb{E}_J(\mathcal{P}_{ST})$, which boils down to the calculation of $\mathbb{E}_J \left( \ev{ H_{ST}}{\psi(\gamma^{\star},\beta^{\star}, \Omega^{\star})} \right)$ with the optimal parameters $\gamma^{\star}$, $\beta^{\star}$ and $\Omega^{\star}$. For this we start by rewriting $H_{ST}$~\eqref{eq:H_ST} in the following form
\begin{equation} \label{eq:H_ST2}
    H_{ST} = \frac{1}{N} \left(  2 \sum_{i<j} J_{ij} \sigma^z_{ij} + \sum_{t} \sum_{\substack{i<j\\i\neq t\\j\neq t}} J_{ij} \sigma^z_{it} \sigma^z_{tj}\right),
\end{equation}
such that the terms are explicitly separated in single-body terms and two-body terms. However, for $p=1$ the gates that are present in the reverse causal cone of the two-body term $\sigma^z_{it} \sigma^z_{tj}$ are independent of $J_{ij}$. Indeed, the coupling strength $J_{ij}$ does not enter in the $p=1$ RCC of $\sigma^z_{it} \sigma^z_{tj}$, which follows from unitary cancellations, originating from the fact that $\comm{\sigma_{it}^x}{\sigma_{ij}^z}=0$ when $t\neq j$ (and that $\comm{\sigma_{it}^z}{\sigma_{ij}^z}=0$, $\forall j$). Therefore, we have that
\begin{equation}
        \mathbb{E}_J \left(J_{ij}  \ev{\sigma^z_{it} \sigma^z_{tj}}{\psi(\gamma,\beta, \Omega)} \right) 
        =  \mathbb{E}_J(J_{ij}) \; \mathbb{E}_J\left( \ev{ \sigma^z_{it} \sigma^z_{tj}}{\psi(\gamma,\beta, \Omega)} \right) =0, 
\end{equation}
as for spin glasses the couplings are taken from a symmetric distribution around zero. This implies that only the local part in Eq.~\eqref{eq:H_ST2} is contributing. These local terms are calculated in the first part of Appendix~\ref{app:calc_p1}. Therefore, 
\begin{equation} \label{eq:EST_p1}
        \mathbb{E}_J \left( \ev{ H_{ST}}{\psi(\gamma,\beta, \Omega)} \right) \approx 
     2\frac{K}{N}\cos^4(2\Omega) \sin(2\beta) \sin(2\gamma) 
\end{equation}
where the approximation means the same as before, i.e. that the topology of all the RCCs is assumed to be the one of the bulk, shown in Fig.~\ref{fig:RCCs}{(A)} for $p=1$ in blue. The optimal angles, minimizing Eq.~\eqref{eq:EST_p1}, are then the trivial angles $\beta^{\star}=-\gamma^{\star}=\pi/4$ and $\Omega^{\star}=0$. Notice that these angles are also optimal when minimizing the energy of the parity-encoded Hamiltonian, see Eq.~\eqref{eq:ex_HPE}, for small constraints strengths $C<2$ and large $N$ (see also Fig.~\ref{fig:angles_C}).  The latter implies that the algorithm does not generate any entanglement, but only performs local rotations. These local rotations map the state $\ket{+}$ back to a product state in the $z$-basis where the state of each qubit $ij$ is set by $J_{ij}$ (in case $J_{ij} \in \{-1,1\}$). Concretely, $J_{ij}=1 \rightarrow \ket{0}$ and $J_{ij}=-1 \rightarrow \ket{1}$). Because of the fact that there are no correlations generated, the algorithm has vanishing performance 
\begin{equation} \label{eq:P_ST_p1}
    \mathbb{E}_J \left( \mathcal{P}_{ST} \right)= \frac{1}{N^{1/2}} - \frac{1}{N^{3/2}} \quad \textrm{for } p=1.
\end{equation}
This result means that the performance of the $p=1$ algorithm unavoidably drops to random sampling as a function of $N$. From this it can already be anticipated that $p$ will necessarily need to grow  with $N$ in order to obtain a better performance.

\subsection{Larger depth $p>1$}

Now, we will derive an upper bound on the performance for general fixed $p$. This upper bound will follow from counting the number of non-vanishing 2-body RCCs. In particular from~\eqref{eq:H_ST2}, it follows that
\begin{equation}\label{eq:UB}
   \mathbb{E}_J \left( \mathcal{P}_{ST} \right) \leq \frac{1}{N^{1/2}} - \frac{1}{N^{3/2}} \quad + \quad \frac{\# \textrm{ of non-vanishing 2-body RCCs}}{N^{5/2}},
\end{equation}
with the number of non-vanishing reverse causal cones depending on $p$. Here the first two terms are equivalent to $2K/N^{5/2}$, which corresponds to the maximal weight of the onsite terms in Eq.~\eqref{eq:H_ST2}. As we have shown in the previous paragraph, this is optimal when $p=1$, as in this case \textit{all} 2-body RCCs do not contribute. 
Here, we will show that for fixed $p$
\begin{equation}\label{eq:UB_fixedp}
   \mathbb{E}_J \left( \mathcal{P}_{ST} \right) \leq \frac{1}{N^{1/2}} + \mathcal{O}(N^{-3/2}) \quad \forall p \textrm{ if }  N > N^{\star}(p),
\end{equation}
hence that non vanishing 2-body RCCs only contribute to the subleading $\mathcal{O}(N^{-3/2})$ corrections.
Concretely, in order to show~\eqref{eq:UB_fixedp}, we will show that the number of non-vanishing 2-body RCCs only grows linearly with $N$ for fixed $p$.

On the other hand if $p$ is taken to be at least of the order of $N$, the number of non-vanishing two-body reverse causal cones will scale as $\mathcal{O}(N^3)$.

For general $p$, a 2-body term ($i\neq t$ and $j \neq t$)
\begin{equation} 
    \mathbb{E}_J \left( J_{ij}  \ev{\sigma^z_{it} \sigma^z_{tj}}{\psi(\bm{\gamma},\bm{\beta}, \bm{\Omega})} \right)
\end{equation}
will not contribute if, as before, $J_{ij}$ is not part of the reverse causal cone of $\sigma^z_{it} \sigma^z_{tj}$. The number of terms for which $J_{ij}$ is part of the reverse causal cone of $\sigma^z_{it} \sigma^z_{tj}$ scales as $Np$ for fixed $t$. This is illustrated in Fig.~\ref{fig:RCCs}{(C)}, where for the qubit line $t=5$ the colored sites $ij$ are inside the reverse causal cone of $\sigma^z_{it} \sigma^z_{tj}$ for a given $p$. Because of this argument the number of non vanishing terms (summing over all $t$'s as in Eq.~\eqref{eq:H_ST2}) scales at most as $N^2p$. Therefore, this is already sufficient to conclude that $\mathbb{E}_J \left( \mathcal{P}_{ST} \right)$ will decrease proportional to $N^{-1/2}$ for fixed $p$.

\begin{figure}
\centering
\includegraphics[width=0.6\textwidth]{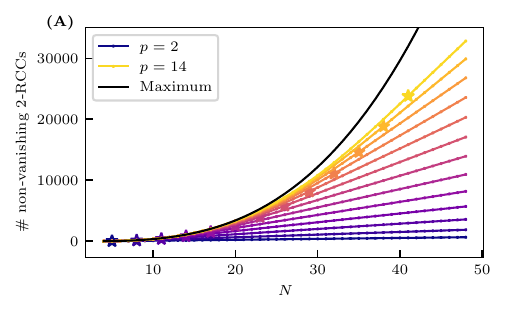}
\includegraphics[width=0.6\textwidth]{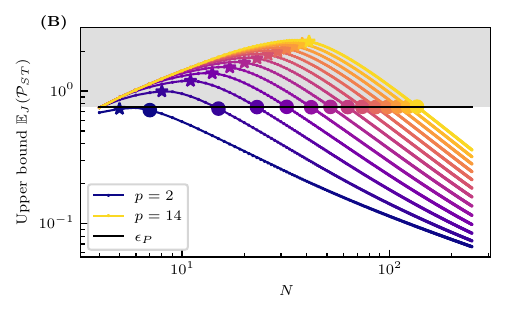} 
\includegraphics[width=0.6\textwidth]{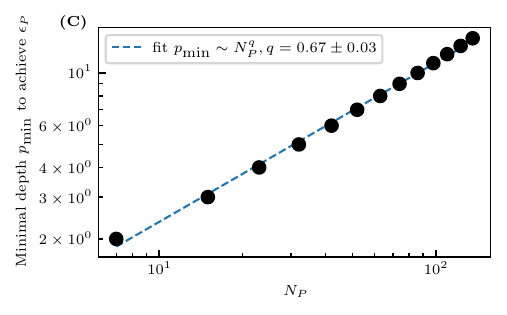} 
\caption{\textbf{(A) Number of non-vanishing two-body reverse causal cones} contributing to the second term on the right-hand side in Eq.~\eqref{eq:H_ST2}. For, fixed $p$, when $N>N^{\star}=3p-1$, the growth becomes exactly linear in $N$. The stars indicate the value of $N^{\star}=3p-1$.  The maximal number, as indicated by the black line $(N-2)K$, is only reached when $p\geq N-1$. This is when each reverse causal cone spans the whole LHZ triangle, as the short base of the triangle contains $N-1$ qubits. \textbf{(B) Upper bound on the performance of parity-encoded QAOA}, as given in the right-hand side of Eq.~\eqref{eq:UB}. As indicated by the grey area, the bounds shown here should only be considered meaningful when they drop below the physical bound, which is the Parisi constant $\epsilon_P$ in the thermodynamic limit. The dots indicate the system size $N_P$ for which the bound is the closest to $\epsilon_P$ for each $p$.  \textbf{(C)} \textbf{The depth $p$ as a function of $N_P$}. We make a numerical fit of the exponent indicating that $p$ should grow with $N$ at least as $N^{0.67}$ to allow for maximal performance.  }
\label{fig:ST_RCC_counting}
\end{figure}

However, we can also fix the prefactor of the decay by realizing that for fixed $p$ actually more 2-body RCCs are vanishing, thus showing that interaction terms do not play a role asymptotically. Even if $J_{ij}$ is part of the reverse causal cone of $\sigma^z_{it} \sigma^z_{tj}$, it is likely (and in fact the bulk contribution, when $N\gg p$) that the reverse causal cones of $\sigma^z_{it}$ and $\sigma^z_{tj}$ are not mutually overlapping. This would imply that the expectation value factorizes
\begin{equation}
    J_{ij} \ev{ \sigma^z_{it} \sigma^z_{tj}}{\psi(\bm{\gamma},\bm{\beta}, \bm{\Omega})}  =  J_{ij} \ev{ \sigma^z_{it}}{\psi(\bm{\gamma},\bm{\beta}, \bm{\Omega})}  \ev{ \sigma^z_{tj}}{\psi(\bm{\gamma},\bm{\beta}, \bm{\Omega})}.
\end{equation}
If this is the case, we can also show that $\mathbb{E}_J \left( J_{ij} \ev{\sigma^z_{it}}{\psi(\bm{\gamma},\bm{\beta}, \bm{\Omega})}\right) = 0$, assuming the RCC of $\sigma^z_{it}$ is dependent on $J_{ij}$. We elaborate on this in Appendix~\ref{app:2body_RCC}.

In Fig.~\ref{fig:ST_RCC_counting}, we show the count of the total number of non-vanishing 2-body reverse causal cones. For fixed $p$, when $N>N^{\star}=3p-1$ the number of those non-vanishing contributions starts growing only linearly with $N$ (the value of $N^{\star}$ is a numerical observation from Fig.~\ref{fig:ST_RCC_counting}{(A)}). Therefore, the two-body terms in Eq.~\eqref{eq:H_ST2} do not even contribute to a prefactor in the decay in the large $N$ limit, but only to the subleading $\mathcal{O}(N^{-3/2})$ corrections in the performance~\eqref{eq:UB_fixedp}. Indeed, in Fig.~\ref{fig:ST_RCC_counting}{(B)} we plot the right-hand side of Eq.~\eqref{eq:UB}, illustrating the vanishing performance with ${N^{-1/2}}$ as $N$ becomes large enough for fixed $p$. In this plot it can also be seen that when $N>N^{\star}=3p-1$ the upper bound starts decreasing, but it only drops below the Parisi constant for $N\geq N_P$ where $N_P$ depends on $p$. 

We comment that the upper bound is based on a counting argument, where all non vanishing terms are counted, and thus all contribute with their maximal individual weight (i.e. 1). When $N$ is roughly the same order as $N^{\star}$, the upper bound is not physical as it is above the Parisi value. This is indicated by the shaded grey area in Fig.~\ref{fig:ST_RCC_counting}{(B)}. Therefore, the bound should only be considered meaningful when it drops below the Parisi value $\epsilon_P$ for larger system sizes.

In Fig.~\ref{fig:ST_RCC_counting}{(C)}, we show the values of $N_P$ for each $p$ and fit the exponent on how $p$ should at least grow with problem size to allow for having maximal performance. From this, the necessary requirement to allow for maximal performance is that $p$ should grow with system size, at least as $N^q$ with $q \gtrsim 0.67$, i.e. as $\Omega(N^{0.67})$. At the same time, we have seen that all terms start contributing only when $p\geq N-1$ [see Fig.~\ref{fig:ST_RCC_counting}{(A)}]. This suggest that it is likely that $p$ will need to grow faster than $N^{0.67}$, e.g. linearly, in order to obtain fixed performance in practice.

\subsection{Scaling of the variance} \label{subsec:variance}

We note that the previous arguments so far only discussed the average over all instances. This does in principle not exclude the existence of an `easier' subset of problem instances for which the performance is non vanishing in the large $N$ limit for fixed $p$. To rule out such non-generic side effects, we would need to prove that the variance of the performance over problem instances is also vanishing. Similarly as for the average, we can construct a vanishing upper bound on the (scaling of the) variance
\begin{equation}
\mathbb{V}\textrm{ar}_J(\mathcal{P}_{ST}) = \mathbb{E}_J (\mathcal{P}_{ST}^2 ) - \left[ \mathbb{E}_J (\mathcal{P}_{ST} ) \right]^2.
\end{equation}
From the previous we already know that the last term is vanishing as $N^{-1}+\mathcal{O}(N^{-2})$, therefore we will focus on the first term. The leading contribution to understand the scaling is
\begin{equation} \label{eq:var_leading_term}
\sum_t \sum_{\substack{i<j\\ i\neq t\\j\neq t}} \sum_{t'} \sum_{\substack{i'<j'\\ i'\neq t'\\j'\neq t'}} \mathbb{E}_J\left(  J_{ij} \ev{ \sigma^z_{it} \sigma^z_{tj}}{\psi}   J_{i'j'} \ev{ \sigma^z_{i't'} \sigma^z_{t'j'}}{\psi}\right)
\end{equation} 
By counting the number of terms, this is upper bounded by $N^{6}$. However, similar as before, the term will be vanishing if the RCCs of $\sigma^z_{it} \sigma^z_{tj}$ and $\sigma^z_{i't'} \sigma^z_{t'j'}$ do not depend on $J_{ij}$ \textit{or} on $J_{i'j'}$. A counting argument, as shown in Fig.~\ref{fig:RCCs}{(C)}, however implies that the number of these non vanishing terms will only grow as $\sim N^4$ for fixed $p$, when $N$ is large enough.

Taking into account other effects such as non-overlapping reverse causal cones of the $\sigma^z$ operators, could lead to more insight in the prefactor. However, this argument is sufficient to conclude that
\begin{equation}
\mathbb{V}\textrm{ar}_J(\mathcal{P}_{ST}) = \mathbb{E}_J (\mathcal{P}_{ST}^2 ) - \left[ \mathbb{E}_J (\mathcal{P}_{ST} ) \right]^2 \sim \frac{1}{N} + \mathcal{O}(N^{-2})
\end{equation}
for fixed $p$ and $N$ large enough.

\subsection{Discussion and comparison with standard QAOA}

\begin{table*}
\centering
\begin{tabular}{lll}
\hline\hline
\textbf{} &
Parity-encoded QAOA & Vanilla QAOA \\
\hline
\textbf{Qubits \hspace{0.4em}} & $N(N-1)/2$ & $ N$ \\
\textbf{Number of gates / QAOA layer\hspace{0.4em} } & $\sim N^2$ & $\sim N^2$ \\
\textbf{Depth / QAOA layer \hspace{0.4em}} & constant & $\sim N$ \\
\textbf{Number of QAOA layers \hspace{0.4em}} & $\sim N^{q}$, $q \gtrsim 0.67$ & constant \\
\textbf{Total number of gates \hspace{0.4em}} & $\sim N^{2+q}$, $q \gtrsim 0.67$ & $\sim N^2$ \\
\textbf{Total depth} & $\sim N^{q},$ $q \gtrsim 0.67$ & $\sim N$ \\
\hline\hline
\end{tabular} 
\caption{Comparison of the scaling with $N$ for the two QAOA variants to solve Sherrington-Kirkpatrick spin-glass models with fixed performance. Vanilla QAOA is the regular version of QAOA for spin-glass models as discussed in Ref.~\cite{Farhi2022}. While for vanilla QAOA the scalings contained in this table are proven, for parity-encoded QAOA they should be interpreted as minimal (necessary) requirements, e.g. the total depth of the parity-encoded algorithm must grow at least as $N^{0.67}$.  }
\label{tab:scalingN}
\end{table*}

From the bounds derived in the beginning of this section, we can conclude that the mapping from the non-local optimization problem to the geometrically local one comes with strong limitations on the performance of local algorithms attempting to solve the problem. Indeed, we derived that the minimum requirement to achieve a fixed performance is to run a non-local algorithm (in the sense that the depth needs to scale as a function of $N$) on the parity-encoded qubits. However, the non-local parity-encoded algorithm needs to act on $K=N(N-1)/2$ qubits. In contrast to vanilla QAOA, which is also a non-local algorithm but which acts on only $N$ qubits, and furthermore has a fixed performance with the number of vanilla QAOA layers~\cite{Farhi2022}.

To summarize our findings, in Table.~\ref{tab:scalingN}, we compare the scaling with $N$ for parity-encoded QAOA with the standard vanilla QAOA to get a fixed performance ratio for solving spin-glass models. As demonstrated in Ref.~\cite{Farhi2022}, the standard QAOA implementation requires only fixed $p$ to get fixed performance in the large $N$ limit. The caveat, however, is that the physical depth of each QAOA layer will scale with $N$~\cite{OGorman2019}. This is because of the assumption, which is true in all currently available hardware platforms, that we cannot perform all two-body gates in parallel. However, for parity-encoded QAOA the physical depth of each layer is constant. On the other hand, as we have discussed previously, the number of layers for parity-encoded QAOA must grow with system size in order to have non vanishing performance. From the upper bound on the performance that is based on the counting argument, we find an exponent of $q \approx 0.67$ to allow for maximal performance [see also Fig.~\ref{fig:RCC_counting}{(C)}]. This formally leaves room for a speedup with respect to vanilla QAOA, in the sense of having the possibility to get the same average performance ratios with a polynomially smaller circuit. Although this is not guaranteed and it would still need to be proven that such performances can be achieved in practice. 

As there is a quadratic qubit overhead in the parity-encoded QAOA, the total number of gates needed for this algorithm scales less favorably than for vanilla QAOA. However, we have not explicitly added noise to the circuit in order to investigate whether there could be some algorithmic robustness under noise. 

We comment that for small system sizes the scalings presented in Table~\ref{tab:scalingN} might not be noticeable as such, because of the presence of finite-size effect. The \texttt{SWAP}-network routing that would be necessary for Vanilla QAOA~\cite{OGorman2019}, adds only a prefactor to the scaling of the depth of a QAOA layer, and therefore does not alter the scaling which is $\mathcal{O}(N)$.

\section{Tensor-network calculations} \label{sec:tn}
In this section, we study the parity-encoded QAOA algorithm numerically, and confirm the predictions from the previous section. We perform exact calculations  by representing the state $\rho(\bm{\gamma},\bm{\beta}, \bm{\Omega}) = \ket{\psi(\bm{\gamma},\bm{\beta}, \bm{\Omega})}\bra{\psi(\bm{\gamma},\bm{\beta}, \bm{\Omega})}$ as a tensor network using the \texttt{QUIMB} package~\cite{Gray2018quimb}. In the tensor-network picture, all the gates are represented by tensors with `input' and `output' indices for each qubit it acts on~\cite{Shi2008}. Then, we calculate each observable that is present in $H_{ST}$ or $H_{PE}$ as a tensor-network contraction over the corresponding reverse causal cone of the Hamiltonian terms. For more details about the tensor-network calculations and for a pseudocode, we refer to Appendix~\ref{app:TN}.

\subsection{Optimized angles for $H_{ST}$} \label{subsec:opt_HST}
We start by optimizing all the QAOA angles on an instance-by-instance basis in order to minimize $\ev{H_{ST}}{\psi(\bm{\gamma},\bm{\beta}, \bm{\Omega})}$, and thus maximize the performance~\eqref{eq:P_ST}. For the optimization, we use the `L-BFGS-B’ implementation of \texttt{SciPy}. This is an optimizer with bounded parameter intervals, therefore we restrict the angle search space to symmetry determined ranges [see Appendix~\ref{app:angle_sym}]. Within these ranges, we initialize the parameters uniformly random.

\begin{figure}
\centering
\includegraphics[width=0.6\textwidth]{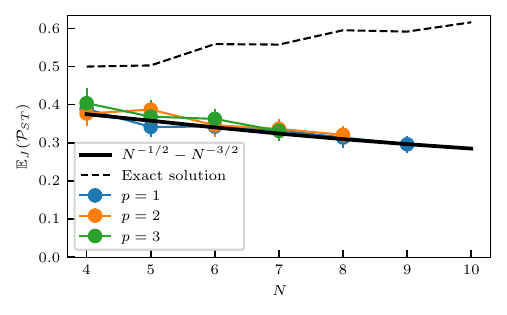}
\caption{\textbf{Numerical performance ratios with the angles optimized instance-by-instance} to minimize the spanning-tree cost function Eq.~\eqref{eq:H_ST}. The averages are taken over 200 instances, and the error bars show three times the standard error of the mean. Histograms over the found angles are shown in Fig.~\ref{fig:Hist_ST}. The performance ratio remains very close to the optimal $p=1$ analytic result [see Eq.~\eqref{eq:P_ST_p1}], indicated by the thick black line, even when increasing $p$. The exact solution is given by the dashed black line. }
\label{fig:P_ST_optimal}
\end{figure}

\begin{figure}
\centering
\includegraphics[width=0.6\textwidth]{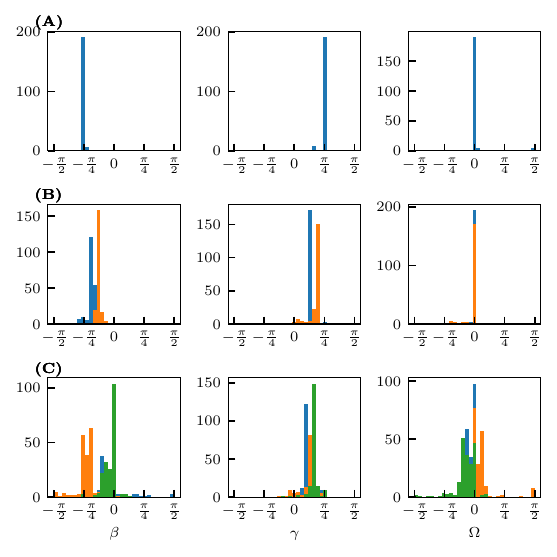}
\caption{\textbf{Histograms over the optimal QAOA angles, minimizing the cost function~\eqref{eq:H_ST} }for $N=7$ and 200 uniformly sampled instances. (A) $p=1$, the angles of the first layer are indicated in blue, (B) $p=2$, angles of the second layer shown in orange,  and (C) $p=3$, angles of the third layer shown in green.  }
\label{fig:Hist_ST}
\end{figure}

In Fig.~\ref{fig:P_ST_optimal}, we plot the numerically determined performance ratio averaged over 200 sampled instances. We see that, in particular for $p=1$, this matches exactly the result~\eqref{eq:P_ST_p1}. However, also for higher $p>1$ it does not improve much beyond this result, consistent with~\eqref{eq:UB_fixedp}. However, also for very small system sizes for which all the RCCs could in principle contribute, we do not obtain better results. This is e.g. the case for $N=4$ and $p=3$. In this figure, we also show the exact solution (dashed black lines) for comparison. In the thermodynamic limit, this converges to the Parisi constant, see Eq.~\eqref{eq:parisi_limit}. However, because of the quadratic qubit overhead of the parity encoding and the underlying structure of the RCCs, we are numerically limited to very small problem sizes. For such small system sizes, there are strong finite size effects. For instance, one such effect is the seemingly different behavior for even and odd system sizes, as is clearly visible in the curve of the exact solution.

In Fig.~\ref{fig:Hist_ST}, we show the histograms over the optimal angles found for $N=7$ at various parity-encoded QAOA depths. Clearly, these angles concentrate around specific values. For $p=1$, they exactly match the predicted angles ($\gamma=-\beta=\pi/4$ and $\Omega=0$) even on an instance-by-instance basis. This is because of self-averaging over the coupling constants $J_{ij}$ that define a problem instance. However, also for $p=3$ the optimizer still finds values for all $\Omega$'s that are near zero for many instances. In those cases, the algorithm does not generate any entanglement, and is thus strictly local. This also explains why the performance on average is not much better than $N^{-1/2}-N^{-3/2}$, which is the best average performance when only local rotations are allowed ($p=1$). 

By Eq.~\eqref{eq:UB_fixedp}, we expect that the performance will drop as in the $p=1$ case for any $p$, as long as $N$ is large enough. (And that therefore all optimal $\Omega$'s will cluster around zero.)  However,  Fig.~\ref{fig:P_ST_optimal} already shows this behavior for the smallest $N$, for which all RCCs could contribute non trivially. This is somewhat surprising, but however suggest that the parameter landscape has a broad (and perhaps only local) minimum for trivial $\Omega$'s, that forms a large basin of attraction for the optimizer.


\subsection{Trotter angles}
In order to check that we can do better than the $p=1$ result in a consistent way, we now take the QAOA angles according to the following Trotter scheme
\begin{align}
\begin{split} \label{eq:Trotter}
    \gamma_i &= \frac{i-1/2}{p} \frac{t_{\mathrm{max}}}{p} \\ 
    \beta_i &= -\left( 1-\frac{i-1/2}{p} \right) \frac{t_{\mathrm{max}}}{p} \\ 
    \Omega_i &= -\frac{i-1/2}{p} \frac{t_{\mathrm{max}}}{p}. \\ 
\end{split}
\end{align}
Similar schemes have been considered for standard QAOA to initialize the optimizers~\cite{Sack2021}. Here, $t_{\mathrm{max}}$ is a parameter that we tuned for each $p$ (and $N$) in order to maximize the spanning-tree performance. However, we found fairly flat curves in the performance when varying this parameter. Hence, we do not believe that the presented results are very sensitive with respect to this parameter. Note that the assumed sign difference between $\gamma_i$ and $\Omega_i$ originates from the fact that unsatisfied constraints give energy penalties [see Eq.~\eqref{eq:H_PC}]. 
 
\begin{figure}
\centering
\includegraphics[width=0.6\textwidth]{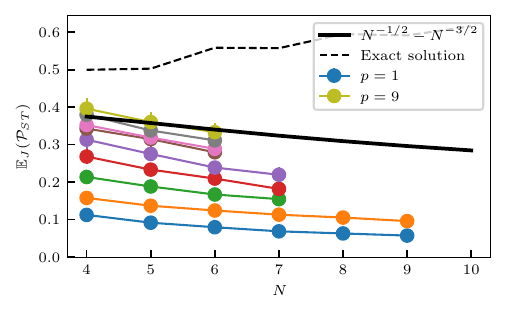}
\caption{\textbf{Numerical performance ratios with angles taken according to the Trotter scheme}~\eqref{eq:Trotter}. The parameter $t_{\mathrm{max}}$ has been roughly optimized to $[0.8,1.4,2.0,2.6,3.2,3.8,4.4,5.0,5.6]$ in units of $J$ for $p=1,2,\dots 9$ and $N=6$ (with slight variations for different $N$). These ratios are compared to (i) the optimal $p=1$ analytical result (thick black line), and (ii) the ratio of the exact solution (dashed black line).  The averages are taken over 200 instances, and the error bars show three times the standard error of the mean.}
\label{fig:P_ST_trotter}
\end{figure}

 In Fig.~\ref{fig:P_ST_trotter}, we show the performance of this scheme averaged over some randomly sampled instances. Of course, as the choices of the angles are not optimal here, the performance lies below the $p=1$ result derived in the previous section [see Eq.~\eqref{eq:P_ST_p1}]. However, this scheme is guaranteed to improve the performance as a function of $p$. We however found that we need to increase to around $p=9$ to see a first improvement for small instances with respect to the $p=1$ algorithm, which is a trivial classical algorithm as it performs only local rotations and generates no entanglement.
 
\subsection{Optimized angles for $H_{PE}$} \label{subsec:opt_HPE}

\begin{figure}
\centering
\includegraphics[width=0.6\textwidth]{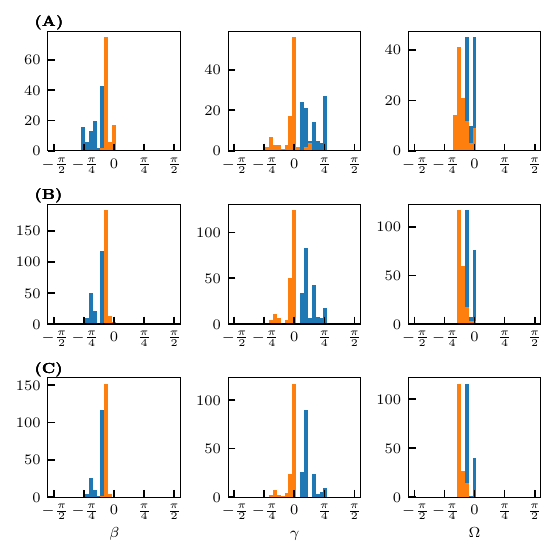}
\caption{\textbf{Histograms over the optimal QAOA angles, minimizing the cost function~\eqref{eq:H_PC}}, for $p = 2$ QAOA layers. The constraint strength is taken $C=3$. (A-C) Logical system sizes $N=6,9,12$ and $N=12$, corresponding respectively to $K=15,36,66$ parity qubits. The peaks grow stronger as the system size increases.}
\label{fig:Hist_PC}
\end{figure}

Finally, we also show some numerically optimized angles that minimize the energy of the Hamiltonian~\eqref{eq:H_PC}, which contains explicit energy penalties for broken constraints. For $p=1$, we can find the optimal angles from minimizing Eq.~\eqref{eq:ex_HPE} [see also Fig.~\ref{fig:angles_C}], or find them on an instance-by-instance basis by numerical evaluation. Again we use the `L-BFGS-B’ algorithm from \texttt{SciPy} for this with the angle intervals restricted as in Appendix~\ref{app:angle_sym} and random initialization. For $p=2$, we show the histograms over the numerically found angles in Fig.~\ref{fig:Hist_PC} for the constraint strength $C=3$. Here, we observe that the clusters of optimal angles grow stronger as the problem size increases, and that some finite-size satellite peaks are disappearing. In particular, the optimal $\Omega_1$, shown in blue, clearly shifts away from 0 to a non-trivial value. This is expected for two reasons. First, for larger system sizes the probability of sampling an `easy' problem (with no, or very few frustration) decreases. Second, contrary to the spanning-tree cost function Eq.~\eqref{eq:H_ST} for which many two-body terms vanish, the multi-body RCCs do not vanish now. This leads to a QAOA circuit that generates entanglement. 

\begin{figure}
\centering
\includegraphics[width=0.6\textwidth]{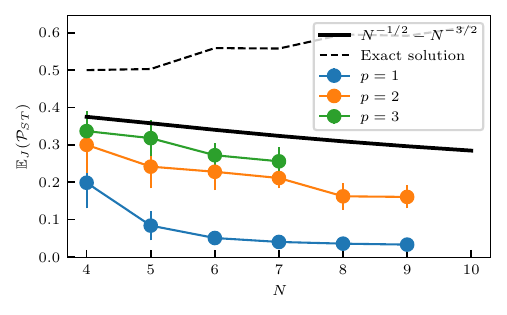}
\caption{\textbf{Numerical performance ratios with the angles optimized instance-by-instance}, to minimize the parity-encoded cost function~\eqref{eq:H_PC}. The constraint strength in this cost function $C=3$. The averages are taken over 100 spin-glass instances, and the errorbars show three times the standard error of the mean.}
\label{fig:P_ST_encoded}
\end{figure}

However, in terms of performance in the spanning tree picture, these angles perform on average worse than $N^{-1/2} - N^{-3/2}$ for small $p$. We show this in Fig.~\ref{fig:P_ST_encoded} up to $p=3$ for small system sizes. For $p=1$, this behavior is expected as the trivial angles are optimal and lead to $N^{-1/2} - N^{-3/2}$ when maximizing the spanning tree performance [see Eq.~\eqref{eq:EST_p1}]. However, because we minimized the energy of Eq.~\eqref{eq:H_PC} with $C=3$, we find non-trivial angles that thus lead to a performance lower than $N^{-1/2} - N^{-3/2}$.  Upon increasing $p$, improvement is guaranteed (see also the previous paragraph).  
However, we still do not improve on $N^{-1/2} - N^{-3/2}$, even for $p=3$. This result shows that after a small number of layers, the state did not satisfy enough constraints, and did not enter in a sufficiently low-energy regime for which the spectra of both Hamiltonians~\eqref{eq:H_PC} and~\eqref{eq:H_ST} become compatible. 

\subsection{Comparison to vanilla QAOA}
For completeness, we also report the results of vanilla QAOA for the same small system sizes. Here we thus optimize the ansatz defined on $N$ qubits, 
\begin{equation}
    \ket{\psi(\bm{\beta}',\bm{\gamma}')} = U_M(\beta_p')U_P(\gamma_p') \dots U_M(\beta_1')U_P(\gamma_1') \ket{+}
\end{equation}
with $U_M(\beta') =  e^{-i\beta'\sum_{i=0}^{N-1} X_i}$, $U_P(\gamma') =  e^{-i\gamma'H_P}$ with $H_P$ given by Eq.~\eqref{eq:SK_log}, and $\ket{+} = \bigotimes_{i=0}^{N-1} \frac{1}{\sqrt{2}}(\ket{0}+\ket{1}) $. In this case, the RCC concept is not very useful as the QAOA, by the problem definition, `sees already the whole graph' at any layer~\cite{Farhi2020,Farhi2020a}. 

\begin{figure}
\centering
\includegraphics[width=0.6\textwidth]{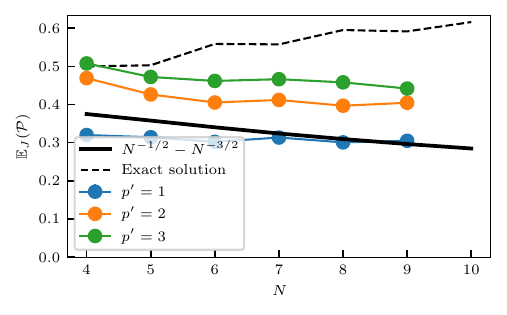}
\caption{\textbf{Vanilla QAOA performance ratios.} These ratios are fixed as $N\rightarrow \infty$, there are only minor finite-size effects.}
\label{fig:P_vanilla}
\end{figure}

In Fig.~\ref{fig:P_vanilla} we show the corresponding performance metric
 \begin{equation} \label{eq:P}
     \mathcal{P}(\bm{\gamma}',\bm{\beta}') = -\frac{\ev{H_{P}}{\psi(\bm{\gamma}',\bm{\beta}')}}{N^{3/2}},
 \end{equation}
 averaged over some instances. These results can be directly compared to the parity results shown in Figs.~\ref{fig:P_ST_optimal},~\ref{fig:P_ST_trotter} and~\ref{fig:P_ST_encoded}. For comparison we also plot $N^{-1/2} - N^{-3/2}$ here. While the parity results in the aforementioned figures struggle to overcome $N^{-1/2} - N^{-3/2}$, the vanilla QAOA performs better than this for $p'=2$ for all $N$ and for $p=1$ for $N\gtrsim 9$. \\
 
Vanilla QAOA has proven performance guarantees for fixed $p'$ in the large $N$ limit. For $p'=1,2,3$ these asymptotic values are respectively $0.30,0.41,0.47$, and can be found in Ref.~\cite{Farhi2022}. This is thus very close to what we found numerically for small system sizes, and allows us to conclude that there are not much finite-size effects in the performance of vanilla QAOA. This should be contrasted with the parity-encoded algorithm for which we have seen that the performance drops asymptotically as $N^{-1/2}$ for any fixed $p$. 

\section{Conclusion and outlook} \label{sec:concl}
In this paper, we have investigated the scaling of parity-encoded QAOA for paradigmatic spin-glass models. To understand the quality of the algorithm, we have calculated the average logical energy of decoding along a qubit-line spanning tree. For fixed depth $p=1$, the performance vanishes exactly as $N^{-1/2} -N^{-3/2}$. By counting the number of non-vanishing 2-body RCCs, we show that the performance is also asymptotically vanishing as $N^{-1/2}$ for arbitrary $p$. This leads to the conclusion that $p$ has to scale up with system size in order to get fixed performance. Our upper bound based on these RCCs counting arguments [see Eq.~\eqref{eq:UB}], sets the requirement that $p$ needs to scale at least as $N^{q}$ with $q\gtrsim 0.67$ in order to allow for maximal performance. 

Formally, this result does not exclude a speed-up of parity-encoded QAOA with respect to vanilla QAOA. This is because the total depth of the algorithm could potentially be more favorable [see Table~\ref{tab:scalingN}]. However, this would require that increasing $p$ actually leads to increased performance. We performed small-scale numerical experiments, and found it in practice hard to obtain better performances than the $p=1$ algorithm [see Figs.~\ref{fig:P_ST_optimal},\ref{fig:P_ST_trotter},\ref{fig:P_ST_encoded}], which performs local rotations and is therefore a trivial and entirely classical algorithm. This follows from the fact that the entangling gate drops out in the parity-encoded QAOA circuit in the $p=1$ case~\eqref{eq:pqaoa_gates}. Therefore we are left with only single-qubit rotations that in this case rotate the parity qubits to the sign of their associated local field $J_{ij}$ (or the interaction between the logical qubits $i$ and $j$). We however find numerically that the parity-encoded algorithm struggles to even outperform this trivial algorithm when $p$ is increased. 

As for all variational algorithms, it is uncertain in the general case that the numerical optimization procedure always yields the globally optimal parameters. However, for some cases we considered here, we could at least approximately compare to analytical formulas (e.g. in the $p=1$ case). In these cases, we found good correspondence between the parameters found by numerical optimization (on an instance-by-instance basis) and the parameters obtained from minimizing formulas that analytically average over all instances. Such a correspondence can also be understood from the fact that spin glasses are `self-averaging' in the sense averaging over the energy over instances is equivalent to averaging different energy contributions within a single `typical' instance. This makes us confident that our numerical results presented in Secs.~\ref{subsec:opt_HST} and~\ref{subsec:opt_HPE} are not dependent on the chosen numerical optimization procedure. 

In this work, we have always considered average performances based on energy expectation values, and we have thus not sampled candidate solutions explicitly. Typically, sampling is also not straightforward from a tensor-network state that is not a matrix-product state. When candidate solutions are however sampled by e.g. state-vector simulation or actual device runs, in principle a post-selection procedure could be applied to only return the candidate solution with the lowest energy. Checking the energy of a candidate solution in the logical picture comes with a classical computational cost of $\mathcal{O}(N^2)$, corresponding to a matrix-vector inner product. Although this may seem like a small computational cost, it is non negligible and, in fact, even overrules the cost scaling (or the depth) of the quantum algorithms considered, see Table~\ref{tab:scalingN}. In the parity-encoded picture, the cost of post-selecting on the best logical solution increases to $\mathcal{O}(N^3)$ because we consider the decoding according to $N$ spanning trees leading to $N$ candidate solutions in the logical picture that need to be checked for every sample from the parity-encoded circuit. We however believe that due to the concentration results of both vanilla QAOA [see Ref.~\cite{Farhi2022}] and parity-encoded QAOA [see~\ref{subsec:variance}], the best solution will not deviate much from the mean, at least for larger system sizes. For small system sizes, there might be finite-size effects due to post-selection that go however beyond the scope of our tensor-network investigation.
We conclude that the parity-encoded QAOA is not showing promising results when compared to vanilla QAOA. 

We leave open if the parity-encoding could work better for different problem classes. However, our work indicates that there is little hope of finding good quality solutions of an originally non-local optimization problem by solving it with a geometrically local reformulation and geometrically local algorithm. We expect that our arguments of RCCs that factorize and eventually drop out, can also be applied on different problems and could also be used to derive performance bounds for other algorithms.

\section{Acknowledgements}
We thank Glen Bigan Mbeng, Michael Fellner, Martin Lanthaler, Maximilian Lutz, Pedro Parrado, PV Sriluckshmy and Anita Weidinger for useful discussions. This project is supported by the Federal Ministry for Economic Affairs and Climate Action on the basis of a decision by the German Bundestag through the project Quantum-enabling Services and Tools for Industrial Applications (QuaST). QuaST aims to facilitate the access to quantum-based solutions for optimization problems.

\newpage
\appendix

\section{Angle symmetries} \label{app:angle_sym}
A angle symmetry $(\bm{\beta},\bm{\gamma},\bm{\Omega})\rightarrow (\bm{\beta}',\bm{\gamma}',\bm{\Omega}')$ in the QAOA algorithm leaves the value of the cost function unchanged, i.e. 
$$E(\bm{\beta},\bm{\gamma},\bm{\Omega}) = E(\bm{\beta}',\bm{\gamma}',\bm{\Omega}') $$ 
with $E(\bm{\beta},\bm{\gamma},\bm{\Omega})  = \ev{H}{\psi(\bm{\beta},\bm{\gamma},\bm{\Omega})}$, where $H$ can be either $H_{PE}$ as in Eq.~\eqref{eq:H_PC}, or $H_{ST}$ as in Eq.~\eqref{eq:H_ST}. 
The most obvious symmetry is a global `time-reversal' symmetry implying that $E(\bm{\beta},\bm{\gamma},\bm{\Omega})  = E(-\bm{\beta},-\bm{\gamma},-\bm{\Omega})$. 

Another very general (and local) QAOA symmetry, arises from the standard form of the mixing Hamiltonian, we have that
\begin{equation} \label{eq:angle_symmetry_beta}
     e^{i(\beta_i+\pi)H_x} = \prod_{i<j} (-\cos(\beta_i) - \sin(\beta_i)\sigma_{ij}^x) = (-1)^{K} e^{i\beta_iH_x}.
\end{equation}

The possible minus sign is irrelevant for the evaluation of the cost function. Therefore, the size of the range of each $\beta_i$ can be restricted to half the unit circle, e.g $(-\pi/2,\pi/2]$. 

Next, the plaquette operator 
\begin{equation}
    H_c = \sum_{i=0}^{N-2} \sum_{j=i+1}^{N-2} \sigma^z_{ij} \sigma^z_{ij+1} \sigma^z_{i+1j+1} [\sigma^z_{i+1j}] 
\end{equation}
has eigenvalues that are either even or odd, depending on whether the number of plaquettes is even or odd. If it is odd, we have that
\begin{equation}
    e^{i(\Omega_i + \pi) H_c} = - e^{i\Omega_i  H_c},
\end{equation}
and if it is even
\begin{equation}
    e^{i(\Omega_i + \pi) H_c} = e^{i\Omega_i  H_c}.   
\end{equation}
Therefore, we can also restrict each $\Omega_i \in (-\pi/2,\pi/2]$. 

Finally, the symmetries on the $\gamma$-angles are problem specific. However with our choice of $ J_{ij} \in \{-1,1\}$, exactly the same argument as in~\eqref{eq:angle_symmetry_beta} applies. So in conclusion the search range for the optimal angles can be bounded to $(-\pi/2,\pi/2]$ for every angle except one, for which the sign can be fixed in addition. 

We finally note that further geometric simplifications could lead to even more restricted ranges. For example, if our original problem would map to a parity configuration that does not simultaneously contain square and triangular plaquettes. Indeed, if it would only contains squares, the plaquette operator would be $\mathbb{Z}_2$ symmetric. While if it would only contain triangles, a global spin flip leads to a global minus sign. This can then in both cases be used to restrict the angles further.

\section{Average expectation values for $p=1$} \label{app:calc_p1}

\subsection{Onsite term}

\begin{figure}
    \centering
    \includegraphics[width=0.45\textwidth]{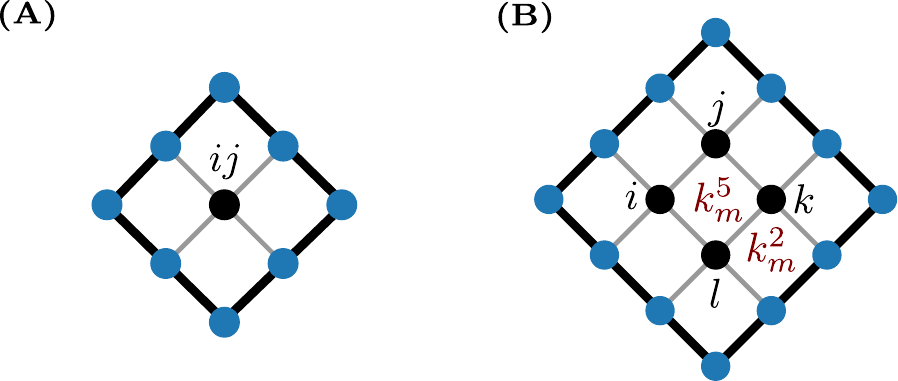}
    \caption{(A) The $p=1$ reverse causal cone of the onsite operator. (B) The $p=1$ reverse causal cone of the plaquette operator. The elements of a constraint vector, e.g. $\bm{k}_m$, should be seen as a product of the elements of a $\bm{z}$-vector as given by the constraints, e.g. $k_m^5 = z_m^i z_m^j z_m^k z_m^l $ (we use single case site labels here). }
    \label{fig:RCC_p1}
\end{figure}

Our goal is to calculate the term $\mathbb{E}_J \ev{J_{ij} \sigma^z_{ij}}{\psi(\gamma_1,\beta_1, \Omega_1)}$. Only the gates that are part of the RCC of the operators $\sigma^z_{ij}$ contribute to this expectation value, and we will only consider the bulk (i.e. square) RCC [see Fig.~\ref{fig:RCC_p1}{(A)}]. We start by inserting identities and by grouping the phases. By $\bm{z}_1,\bm{z}_{-1},\bm{z}_m$ we denote bitstrings of length $(2p+1)^2=9$ describing a configuration on the RCC, $\bm{z}_1^{ij} \in \{-1,+1\}$, and by $\sum_{\{\bm{z}\}}$ we mean summing over all $(2^{3\times9})$ variable configurations.

\begin{align}
    \mathbb{E}_J &\left( \ev{J_{ij} \sigma^z_{ij}}{\psi(\gamma_1,\beta_1, \Omega_1)} \right) \\
    &= \mathbb{E}_J \left(\ev{e^{i J_{ij}\gamma_1 \sigma^z_{ij}}  e^{i \Omega_1 H_c} e^{i\beta_1 \sigma^x_{ij}}  J_{ij} \sigma^z_{ij} e^{-i\beta_1 \sigma^x_{ij}}  e^{-i J_{ij}\Omega_1 H_c} e^{-i J_{ij}\gamma_1 \sigma^z_{ij}}   }{x}  \right) \\
    &= \frac{1}{2^9} \mathbb{E}_J ( \sum_{\{\bm{z}\}} \ev{e^{i J_{ij}\gamma_1 \sigma^z_{ij}} }{\bm{z}_1} \ev{e^{i \Omega_1 H_c}}{\bm{z}_1} \mel{\bm{z}_1}{e^{i\beta_1 \sigma^x_{ij}} }{\bm{z}_m}  
    \ev{J_{ij} \sigma^z_{ij}}{\bm{z}_m} \mel{\bm{z}_m}{e^{-i\beta_1 \sigma^x_{ij}} }{\bm{z}_{-1}}  
    \\ &\quad \quad \quad \quad \quad \ev{-e^{i \Omega_1 H_c}}{\bm{z}_{-1}} \ev{e^{-i J_{ij}\gamma_1 \sigma^z_{ij}} }{\bm{z}_{-1}} ) \\
    &= \frac{1}{2^9}  \sum_{\{\bm{z}\}} e^{i \Omega_1 (H_c(\bm{z}_1) - H_c(\bm{z}_{-1}))} \mel{\bm{z}_1}{e^{i\beta_1 \sigma^x_{ij}} }{\bm{z}_m}  
    \mel{\bm{z}_m}{e^{-i\beta_1 \sigma^x_{ij}} }{\bm{z}_{-1}}  
    \mathbb{E}_J \left( J_{ij} z_m^{ij} e^{i J_{ij}\gamma_1 (z_1^{ij}-z_{-1}^{ij})}  \right) \\
\end{align}

At this point, we can easily perform the average $ \mathbb{E}_J$, and in a second step redefine $\bm{z}_1 \rightarrow \bm{z}_1\bm{z}_m$, and $\bm{z}_{-1} \rightarrow \bm{z}_{-1}\bm{z}_m$. 

\begin{align}
    \mathbb{E}_J &\left( \ev{J_{ij} \sigma^z_{ij}}{\psi(\gamma_1,\beta_1, \Omega_1)} \right) \\
    &= \frac{1}{2^9}  \sum_{\{\bm{z}\}} e^{i \Omega_1 (H_c(\bm{z}_1) - H_c(\bm{z}_{-1}))} \mel{\bm{z}_1}{e^{i\beta_1 \sigma^x_{ij}} }{\bm{z}_m}  
    \mel{\bm{z}_m}{e^{-i\beta_1 \sigma^x_{ij}} }{\bm{z}_{-1}}  
    i z_m^{ij} \sin(\gamma_1 (z_{1}^{ij}-z_{-1}^{ij}) )  \\
     &= \frac{1}{2^9}  \sum_{\{\bm{z}\}} e^{i \Omega_1 (H_c(\bm{z}_1\bm{z}_m) - H_c(\bm{z}_{-1}\bm{z}_m))} \mel{\bm{z}_1}{e^{i\beta_1 \sigma^x_{ij}} }{\bm{1}}  
    \mel{\bm{1}}{e^{-i\beta_1 \sigma^x_{ij}} }{\bm{z}_{-1}}  
    i z_m^{ij} \sin(\gamma_1 (z_{1}^{ij}-z_{-1}^{ij})z_m^{ij} )  \\
\end{align}

Where we used the fact that $\mel{\bm{z}_1 \bm{z}_m}{e^{i\beta \sigma^x_{ij}} }{\bm{z}_m} = \mel{\bm{z}_1}{e^{i\beta \sigma^x_{ij}} }{\bm{1}}$. Therefore all the vector elements of $\bm{z}_1$ and $\bm{z}_{-1}$ must be $1$, except on position $(ij)$. The dependence on $z_m^{ij}$ vanishes in the last part, such that only the constraints are dependent on the $\bm{z}_m$ bitstrings.

\begin{align}
    \mathbb{E}_J &\left( \ev{J_{ij} \sigma^z_{ij}}{\psi(\gamma_1,\beta_1, \Omega_1)} \right) \\
     &= \frac{1}{2^9}  \sum_{\bm{z}_1,\bm{z}_{-1}} \left( \sum_{\bm{z}_m} e^{i \Omega_1 (H_c(\bm{z}_1\bm{z}_m) - H_c(\bm{z}_{-1}\bm{z}_m))} \right) \mel{\bm{z}_1}{e^{i\beta_1 \sigma^x_{ij}} }{\bm{1}}  
    \mel{\bm{1}}{e^{-i\beta_1 \sigma^x_{ij}} }{\bm{z}_{-1}}  
    i\sin(\gamma_1 (z_{1}^{ij}-z_{-1}^{ij}))  \\
\end{align}

Now we can perform the summation over $\bm{z}_m$. For this, we can go to the eigenbasis of the constraint Hamiltonian. We will denote $\bm{k}_1$ as the constraint bitstring (in this case of length 4) corresponding to $\bm{z}_1$, i.e. $k_1^i = 1$ ($-1$) means that the $i$th constraint is satisfied (broken).
\begin{align}
    \sum_{\bm{z}_m} e^{i \Omega_1 (H_c(\bm{z}_1\bm{z}_m) - H_c(\bm{z}_{-1}\bm{z}_m))} 
    &= 2^{5}  \sum_{\bm{k}_m} \exp(i\Omega_1 (z^{ij}_1-z^{ij}_{-1}) \sum_{i=0}^3 k^i_m ) \\
   &= 2^{5}  \sum_{\bm{k}_m} \prod_{i=0}^3 \exp(i\Omega_1 (z^{ij}_1-z^{ij}_{-1})  k^i_m ) \\
     &= 2^{9} \cos^4(\Omega_1 (z^{ij}_1-z^{ij}_{-1}) ) \\
\end{align}

Plugging it back and performing the remaining summations yields, 

\begin{align}
    \mathbb{E}_J &\left( \ev{J_{ij} \sigma^z_{ij}}{\psi(\gamma_1,\beta_1, \Omega_1)} \right) \\
     &= \sum_{\bm{z}_1,\bm{z}_{-1}} \cos^4(\Omega_1 (z^{ij}_1-z^{ij}_{-1}) ) \mel{\bm{z}_1}{e^{i\beta_1 \sigma^x_{ij}} }{\bm{1}}  
    \mel{\bm{1}}{e^{-i\beta_1 \sigma^x_{ij}} }{\bm{z}_{-1}}  
    i\sin(\gamma_1 (z_{1}^{ij}-z_{-1}^{ij}))  \\
    &= \cos^4(2\Omega_1) \sin(2\beta_1) \sin(2\gamma_1)
\end{align}

\subsection{Plaquette term}
Here the goal is to evaluate the term $\mathbb{E}_J \ev{ \sigma^z_{i}\sigma^z_{l}\sigma^z_{k}\sigma^z_{l}}{\psi(\gamma_1,\beta_1, \Omega_1)}$. In order to make the notation slightly lighter, we will use single case labels here [see Fig.~\ref{fig:RCC_p1}{(B)}]. Again, only the gates that are part of the RCC of $\sigma^z_{i}\sigma^z_{j}\sigma^z_{k}\sigma^z_{l}$ contribute to this expectation value. We start by performing similar steps as before, now the bitstrings $\bm{z}_1,\bm{z}_{-1},\bm{z}_m$ have length $(2p+2)^2=16$.
\begin{align}
    \mathbb{E}_J &\left( \ev{ \sigma^z_{i}\sigma^z_{j}\sigma^z_{k}\sigma^z_{l}}{\psi(\gamma_1,\beta_1, \Omega_1)} \right) \\
    &= \mathbb{E}_J \left(\ev{e^{i J_{ij}\gamma_1 \sum_{n\in \square} \sigma^z_{n}}  e^{i \Omega_1 H_c} e^{i\beta_1 \sum_{n\in \square} \sigma^x_{n}}  \sigma^z_{i}\sigma^z_{l}\sigma^z_{k}\sigma^z_{l} e^{-i\beta_1 \sum_{n\in \square} \sigma^x_{n}}  e^{-i J_{ij}\Omega_1 H_c} e^{-i J_{ij}\gamma_1 \sum_{n\in \square} \sigma^z_{n}}   }{x}  \right) \\
    &= \frac{1}{2^{16}} \mathbb{E}_J ( \sum_{\{\bm{z}\}} \ev{e^{i J_{ij}\gamma_1 \sum_{n\in \square} \sigma^z_{n}} }{\bm{z}_1} \ev{e^{i \Omega_1 H_c}}{\bm{z}_1} \mel{\bm{z}_1}{e^{i\beta_1 \sum_{n\in \square} \sigma^x_{n}} }{\bm{z}_m}  
    \ev{\sigma^z_{i}\sigma^z_{j}\sigma^z_{k}\sigma^z_{l}}{\bm{z}_m}
    \\ &\quad \quad \quad \quad \quad 
    \mel{\bm{z}_m}{e^{-i\beta_1 \sum_{n\in \square} \sigma^x_{n}} }{\bm{z}_{-1}}  
    \ev{-e^{i \Omega_1 H_c}}{\bm{z}_{-1}} \ev{e^{-i J_{ij}\gamma_1 \sum_{n\in \square} \sigma^z_{n}} }{\bm{z}_{-1}} ) \\
    &= \frac{1}{2^{16}}  \sum_{\{\bm{z}\}} z_m^i z_m^j z_m^k z_m^l e^{i \Omega_1 (H_c(\bm{z}_1) - H_c(\bm{z}_{-1}))} \mel{\bm{z}_1}{e^{i\beta_1 \sum_{n\in \square} \sigma^x_{n}} }{\bm{z}_m}  
    \mel{\bm{z}_m}{e^{-i\beta_1 \sum_{n\in \square} \sigma^x_{n}} }{\bm{z}_{-1}}  
    \\ &\quad \quad \quad \quad \quad 
    \mathbb{E}_J \left( e^{i \gamma_1 \sum_{n\in \square} J_n (z_1^{n}-z_{-1}^{n})}  \right) \\
     &=\frac{1}{2^{16}}  \sum_{\{\bm{z}\}} z_m^i z_m^j z_m^k z_m^l e^{i \Omega_1 (H_c(\bm{z}_1) - H_c(\bm{z}_{-1}))} \mel{\bm{z}_1}{e^{i\beta_1 \sum_{n\in \square} \sigma^x_{n}} }{\bm{z}_m}  
    \mel{\bm{z}_m}{e^{-i\beta_1 \sum_{n\in \square} \sigma^x_{n}} }{\bm{z}_{-1}}  
    \\ &\quad \quad \quad \quad \quad 
     \prod_{n\in \square} \cos( \gamma_1 (z_1^{n}-z_{-1}^{n})) \\
      &=\frac{1}{2^{16}}  \sum_{\bm{z}_1,\bm{z}_{-1}} \left( \sum_{\bm{z}_m} z_m^i z_m^j z_m^k z_m^l e^{i \Omega_1 (H_c(\bm{z}_1\bm{z}_m) - H_c(\bm{z}_{-1}\bm{z}_m))} \right)
      \mel{\bm{z}_1}{e^{i\beta_1 \sum_{n\in \square} \sigma^x_{n}} }{\bm{1}}  
    \mel{\bm{1}}{e^{-i\beta_1 \sum_{n\in \square} \sigma^x_{n}} }{\bm{z}_{-1}}  
    \\ &\quad \quad \quad \quad \quad 
     \prod_{n\in \square} \cos( \gamma_1 (z_1^{n}-z_{-1}^{n})) 
\end{align}
Now we can again perform the summation over $\bm{z}_m$ first, we have that $z_m^i z_m^j z_m^k z_m^l = k_m^5 $
\begin{align}
    \sum_{\bm{z}_m} z_m^i z_m^j z_m^k z_m^l e^{i \Omega_1 (H_c(\bm{z}_1\bm{z}_m) - H_c(\bm{z}_{-1}\bm{z}_m))}
    &= 2^7 \sum_{\bm{k}_m} k_m^5  e^{i \Omega_1 \sum_{n=0}^8( k_1^n -k_{-1}^n) k_m^n } \\
    &= 2^7 \sum_{\bm{k}_m} \prod_{n=0}^8 k_m^5  e^{i \Omega_1 ( k_1^n -k_{-1}^n) k_m^n } \\
    &= 2^{16} i\sin(\Omega_1(k_1^5 -k_{-1}^5)) \prod_{n\neq 5} \cos(\Omega_1(k_1^n -k_{-1}^n))
\end{align}
Therefore we have that 
\begin{align}
    \mathbb{E}_J &\left( \ev{ \sigma^z_{i}\sigma^z_{l}\sigma^z_{k}\sigma^z_{l}}{\psi(\gamma_1,\beta_1, \Omega_1)} \right) \\
      &= \sum_{\bm{z}_1,\bm{z}_{-1}} i\sin(\Omega_1(k_1^5 -k_{-1}^5)) \prod_{n\neq 5} \cos(\Omega_1(k_1^n -k_{-1}^n))
      \mel{\bm{z}_1}{e^{i\beta_1 \sum_{n\in \square} \sigma^x_{n}} }{\bm{1}}  
    \mel{\bm{1}}{e^{-i\beta_1 \sum_{n\in \square} \sigma^x_{n}} }{\bm{z}_{-1}}  
    \\ &\quad \quad \quad \quad \quad 
     \prod_{n\in \square} \cos( \gamma_1 (z_1^{n}-z_{-1}^{n})) 
\end{align}
We see that $\bm{z}_1$ and $\bm{z}_{-1}$ run each over $2^4=16$ spin configurations (256 configurations in total). However, the $\sin(\Omega_1(k_1^5 -k_{-1}^5))$ implies that an even (odd) number of flips in $\bm{z}_1$ must combine with an odd (even) number of flips in $\bm{z}_{-1}$, to be non vanishing. We list the non-vanishing contribution is the table below.
\begin{table}[htbp]
\begin{center}
\label{tb:addlabel}
\begin{tabular}{llll}
\hline\hline
$\#$ of flips in $\bm{z}_1$ & $\#$ of flips in $\bm{z}_{-1}$ & Multiplicity & Term\\
\hline
0&1&4&$\cos^7{\beta_1} \sin{\beta_1} \cos(2\gamma_1) \sin(2\Omega_1) \cos^3(2\Omega_1)$  \\
0&3&4&$-\cos^5{\beta_1} \sin^3{\beta_1} \cos^3(2\gamma_1) \sin(2\Omega_1) \cos^5(2\Omega_1)$  \\
1&2&12&$ -\cos^5{\beta_1} \sin^3{\beta_1} \cos(2\gamma_1) \sin(2\Omega_1) \cos^3(2\Omega_1)$  \\
1&2&12&$-\cos^5{\beta_1} \sin^3{\beta_1} \cos^3(2\gamma_1) \sin(2\Omega_1) \cos^5(2\Omega_1)$  \\
1&4&4&$ \cos^3{\beta_1} \sin^5{\beta_1} \cos^3(2\gamma_1) \sin(2\Omega_1) \cos^5(2\Omega_1)$  \\
2&3&12&$ \cos^3{\beta_1} \sin^5{\beta_1} \cos^3(2\gamma_1) \sin(2\Omega_1) \cos^5(2\Omega_1)$  \\
2&3&12&$ \cos^3{\beta_1} \sin^5{\beta_1} \cos(2\gamma_1) \sin(2\Omega_1) \cos^3(2\Omega_1)$  \\
3&4&4&$ -\cos{\beta_1} \sin^7{\beta_1} \cos(2\gamma_1) \sin(2\Omega_1) \cos^3(2\Omega_1)$  \\
\hline\hline
\end{tabular}
\end{center}
\caption{Non-vanishing combinations. The terms are symmetric upon interchanging  $\bm{z}_1 \leftrightarrow \bm{z}_{-1}$.}
\end{table}

First, combining the terms with $\cos(2\gamma_1) \cos^3(2\Omega_1)$ gives
\begin{align}
    & 2 \cdot 4 \cos(2\gamma_1)\sin(2\Omega_1) \cos^3(2\Omega_1) \left( \cos^7{\beta_1} \sin{\beta_1} -3\cos^5{\beta_1} \sin^3{\beta_1} + 3\cos^3{\beta_1} \sin^5{\beta_1} -\cos{\beta_1} \sin^7{\beta_1}\right) \\
    &= 4   \cos(2\gamma_1)\sin(2\Omega_1) \cos^3(2\Omega_1) \sin(2\beta_1) \left( \cos^6{\beta_1}  -3\cos^4{\beta_1} \sin^2{\beta_1} + 3\cos^2{\beta_1} \sin^4{\beta_1} - \sin^6{\beta_1}\right)\\
     &= 4   \cos(2\gamma_1)\sin(2\Omega_1) \cos^3(2\Omega_1) \sin(2\beta_1) \cos^3(2\beta_1).
\end{align}
Secondly, combining the terms with $\cos^3(2\gamma_1) \cos^5(2\Omega_1)$ gives
\begin{align}
    & 2 \cdot 16 \cos^3(2\gamma_1) \sin(2\Omega_1) \cos^5(2\Omega_1) \left( \cos^3{\beta_1} \sin^5{\beta_1} - \cos^5{\beta_1} \sin^3{\beta_1} \right) \\ 
    & = -4 \cos^3(2\gamma_1) \sin(2\Omega_1) \cos^5(2\Omega_1) \sin^3(2\beta_1) \cos(2\beta_1)
\end{align}
Combining both finally gives
\begin{align}
    \mathbb{E}_J &\left( \ev{ \sigma^z_{i}\sigma^z_{l}\sigma^z_{k}\sigma^z_{l}}{\psi(\gamma_1,\beta_1, \Omega_1)} \right) \\
      &= 4 \cos(2\gamma_1) \sin(2\Omega_1) \cos^3(2\Omega_1) \sin(2\beta_1) \cos(2\beta_1) \left( \cos^2(2\beta_1) - \cos^2(2\gamma_1) \cos^2(2\Omega_1) \sin^2(2\beta_1) \right)
\end{align}

\subsection{Result $p=1$}
Combining the two previous sections yields
\begin{align}
    \mathbb{E}_J &\left( \ev{ H_{PE}}{\psi(\gamma_1,\beta_1, \Omega_1)} \right) \\
    &\approx K\cos^4(2\Omega_1) \sin(2\beta_1) \sin(2\gamma_1) \\
    &\quad \quad - 4\cdot P \cdot C \cos(2\gamma_1) \sin(2\Omega_1) \cos^3(2\Omega_1) \sin(2\beta_1) \cos(2\beta_1) \left( \cos^2(2\beta_1) - \cos^2(2\gamma_1) \cos^2(2\Omega_1) \sin^2(2\beta_1) \right).
\end{align}
Where we have assumed that all RCC are square. One could easily repeat the calculations for different RCC topologies, but even for small $N \gtrsim 6$, this approximation turns out to perform well. In the large $N$ limit ($N=800$), and for $C=3$, we find the following optimal angles
\begin{align}
    \beta_1 &= 0.219 \\
    \gamma_1 &= -0.167 \\
    \Omega_1 &= 0.263.
\end{align}

\section{Vanishing contribution when the 2-body RCCs of $H_{ST}$ do not overlap} \label{app:2body_RCC}
\begin{figure}
    \centering
    \includegraphics[width=0.99\textwidth]{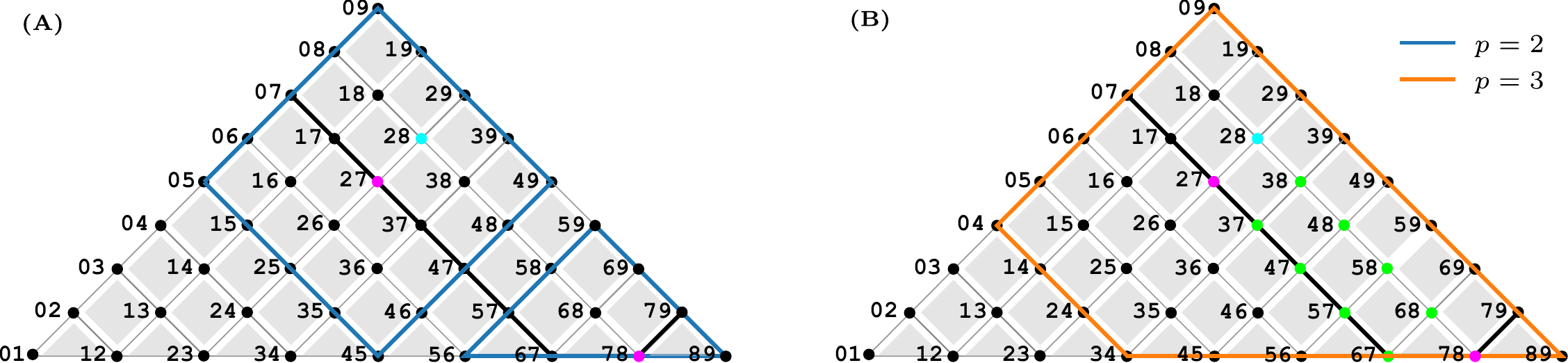}
    \caption{ (A) Sketch of the LHZ triangle for $N=10$ with the $t=7$ qubit line indicated. The boundary of the $p=2$ RCC of $\ev{ \sigma^z_{27} \sigma^z_{78}}{\psi(\bm{\gamma},\bm{\beta}, \bm{\Omega})}$ is indicated in blue. These RCC consist of two isolated parts. Therefore the expectation value factorizes. The expectation value $\ev{ \sigma^z_{27} }{\psi(\bm{\gamma},\bm{\beta}, \bm{\Omega})}$ depends on $J_{28}$. However Eq.~\eqref{eq:example_vanish} is zero because $z_m^{27} z_m^{28}$ cannot be completed to a product of constraints (inside the RCC). (B) When $p=3$, $\ev{ \sigma^z_{27} \sigma^z_{78}}{\psi(\bm{\gamma},\bm{\beta}, \bm{\Omega})}$ does not factorize and the product $z_m^{27} z_m^{28} z_m^{78}$ can be completed as indicated by the path of green dots. }
    \label{fig:example_non_overlapping_RCCs}
\end{figure}

In this appendix, we show why
\begin{equation}
    \mathbb{E}_J \left( J_{ij} \ev{ \sigma^z_{it}}{\psi(\bm{\gamma},\bm{\beta}, \bm{\Omega})}   \ev{ \sigma^z_{tj}}{\psi(\bm{\gamma},\bm{\beta}, \bm{\Omega})} \right)  = 0
\end{equation}
where we assumed that the RCCs of $\sigma^z_{it}$ and $\sigma^z_{tj}$ do not overlap, i.e. that
\begin{equation}
    J_{ij} \ev{ \sigma^z_{it} \sigma^z_{tj}}{\psi(\bm{\gamma},\bm{\beta}, \bm{\Omega})}  = J_{ij} \ev{ \sigma^z_{it}}{\psi(\bm{\gamma},\bm{\beta}, \bm{\Omega})}   \ev{ \sigma^z_{tj}}{\psi(\bm{\gamma},\bm{\beta}, \bm{\Omega})}.
\end{equation}
Additionally, we also assume the $it$ and $tj$ are chosen such that either $ \ev{ \sigma^z_{it}}{\psi(\bm{\gamma},\bm{\beta}, \bm{\Omega})}$ or $ \ev{ \sigma^z_{tj}}{\psi(\bm{\gamma},\bm{\beta}, \bm{\Omega})}$ is dependent on $J_{ij}$. Otherwise we would anyway have a vanishing contribution, simply because $\mathbb{E}_J(J_{ij})=0$. An example of such a case is sketched in Fig.~\ref{fig:example_non_overlapping_RCCs}{(A)}. Here $t=7$, $i=2$ and $j=8$. When $p=2$ the RCCs do not overlap, but $ \ev{ \sigma^z_{27}}{\psi(\bm{\gamma},\bm{\beta}, \bm{\Omega})}$ depends on $J_{28}$. Therefore
\begin{equation} \label{eq:example_vanish}
     \mathbb{E}_J \left( J_{28} \ev{ \sigma^z_{27}}{\psi(\bm{\gamma},\bm{\beta}, \bm{\Omega})}  \right) 
\end{equation}
does not factorize. We can however realize that this still vanishes. Making a similar basis expansion as in the previous section, leads to
\begin{align}
     \mathbb{E}_J & \left( J_{28} \ev{ \sigma^z_{27}}{\psi(\bm{\gamma},\bm{\beta}, \bm{\Omega})}  \right) \\
     &\sim  \sum_{\{\bm{z}\}}z_m^{27} e^{i\Omega_1(H_c(\bm{z}_1\bm{z}_2\bm{z}_m)-H_c(\bm{z}_{-1}\bm{z}_{-2}\bm{z}_m))} e^{i\Omega_2(H_c(\bm{z}_1\bm{z}_m)-H_c(\bm{z}_{-1}\bm{z}_m))}  \sin(\gamma_1(z_1^{28}z_2^{28}z_m^{28}-z_{-1}^{28}z_{-2}^{28}z_m^{28})) \\ & \quad \quad \quad \times G(\gamma_1,\gamma_2,\beta_1,\beta_2,\bm{z}_1,\bm{z}_{-1},\bm{z}_{2},\bm{z}_{-2})\\
     &=  \sum_{\bm{z}_1,\bm{z}_{-1},\bm{z}_{2},\bm{z}_{-2}} \left[\sum_{\bm{z}_m}  z_m^{27} z_m^{28} e^{i\Omega_1(H_c(\bm{z}_1\bm{z}_2\bm{z}_m)-H_c(\bm{z}_{-1}\bm{z}_{-2}\bm{z}_m))} e^{i\Omega_2(H_c(\bm{z}_1\bm{z}_m)-H_c(\bm{z}_{-1}\bm{z}_m))}  \right]  \\ & \quad \quad \quad \times G(\gamma_1,\gamma_2,\beta_1,\beta_2,\bm{z}_1,\bm{z}_{-1},\bm{z}_{2},\bm{z}_{-2}) \sin(\gamma_1(z_1^{28}z_2^{28}-z_{-1}^{28}z_{-2}^{2})),
\end{align}
here $G(\dots)$ should just be read as being a function of many variables, but that crucially has no dependence on $\bm{z}_m$. Now we can see that the factor in square brackets is always zero. To work out $H_c(\bm{z}_1\bm{z}_2\bm{z}_m)$, we would typically go to the eigenbasis of this Hamiltonian. (This has been described by the $\bm{k}$ vectors before.) However, we can never complete the product of $z_m^{27} z_m^{28}$ into a product of $k_m$'s, e.g. $k_m^{27}=z_m^{27}z_m^{28}z_m^{38}z_m^{38}$ (we use the convention that a plaquette is labelled by its leftmost qubit). Therefore each $k_m^{ab}$ that appears in the exponents comes with $z_m^{27} z_m^{28}=1$ in half of the cases, and $z_m^{27} z_m^{28}=-1$ in the other half of the cases. This then sums to zero. This is true for any $p$, as long as these RCCs do not overlap. 

In case the RCCs overlap, as is the case for $p=3$ in our example [see Fig.~\ref{fig:example_non_overlapping_RCCs}{(B)}]. Here, we see that the product $z_m^{27}z_m^{28}z_m^{78}$ can be completed, as indicated by the green dots in Fig.~\ref{fig:example_non_overlapping_RCCs}{(B)}. This then implies that the factor will not be vanishing. Concretely, for our example we would have that
\begin{align}
z_m^{27}z_m^{28}z_m^{78} 
&= z_m^{27} z_m^{28}  \quad z_m^{37}z_m^{37} \quad z_m^{38}z_m^{38} \quad  z_m^{47}z_m^{47}\quad z_m^{48}z_m^{48} \quad z_m^{57}z_m^{57}\quad z_m^{58}z_m^{58} \quad z_m^{67}z_m^{67} \quad z_m^{68}z_m^{68} \quad z_m^{78} \\
&=  z_m^{27} z_m^{28}  z_m^{38} z_m^{37} \quad z_m^{37} z_m^{38}  z_m^{48} z_m^{47} \quad z_m^{47} z_m^{48}  z_m^{58} z_m^{57} \quad z_m^{57} z_m^{58}  z_m^{68} z_m^{67} \quad z_m^{67} z_m^{68}  z_m^{78} \\
&= k_m^{27} \quad k_m^{37} \quad  k_m^{47} \quad  k_m^{57} \quad  k_m^{67}.
\end{align}

\section{Tensor-network calculations and reverse causal cones} \label{app:TN}

\begin{figure}
    \centering
    \includegraphics[width=0.6\textwidth]{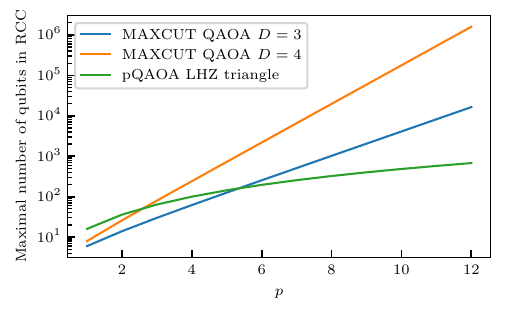}
    \caption{The maximal number of qubits that occur in the RCCs for different QAOA algorithms. Importantly, the contraction complexity does not scale with these numbers for Maxcut QAOA, as the structure of the RCCs here are mostly trees. On the other hand, for parity-encoded QAOA, the structure of the RCCs is square lattice. Therefore, the contraction complexity does scale with quadratic qubit count in this case.  }
    \label{fig:RCC_counting}
\end{figure}

In this appendix, we discuss further details about the tensor-network calculations performed in Sec.~\ref{sec:tn} of the main text. All the calculations presented in the main text are exact, thus we did not perform any truncation in the ``bond’’ dimensions of the tensor network. For these calculations we have used the \texttt{QUIMB} package~\cite{Gray2018quimb}, which allowed for an implementation of the parity-encoded QAOA circuit~\eqref{eq:pqaoa_gates} and the computation of expectation values on the RCCs. We provide pseudocode in Algorithm~\ref{alg:QAOA}. To reproduce our parity-encoded results, one needs to set up the problem graph as the triangle [see e.g. Fig.~\ref{fig:lhz_mapping}], the Hamiltonian as Eq.~\eqref{eq:H_PC} or Eq.~\eqref{eq:H_ST} and the QAOA ansatz as Eq.~\eqref{eq:psi}. Then the angle optimization is carried out in order to minimize the energy of the chosen Hamiltonian. As described in the main text, we have used `L-BFGS-B’ from \texttt{SciPy} for this optimization where we restrict the parameter ranges to the ones of Appendix~\ref{app:angle_sym}. To evaluate local expectation values (i.e. individual terms in the Hamiltonians), \texttt{QUIMB} only contracts tensors (gates) that are within the reverse causal cone of the operators. So the function \textit{LocalExpectationValue} of our pseudocode is already native in \texttt{QUIMB}.

The advantage of the tensor-network picture is that it allows to calculate local expectation values of local quantum circuits for a larger number of qubits than a naive state-vector calculation, on the condition that the depth $p$ is small.  This follows from the fact that each local term that is present in the Hamiltonian as a tensor-network contraction over the tensors that are present in the lightcone (or RCC) of this term. The cost of this procedure in general is dependent on the structure of the RCCs, and on the exact contraction path that is found. The disadvantage of the tensor-network calculation compared to a state-vector simulation is that it is less straightforward to sample. However, in this paper we are only interested in computing expectation values.

\begin{algorithm}
\begin{small}
   \caption{Parity-encoded QAOA}
   \label{alg:QAOA}
   \KwInput{$N$, $p$}

Interactions = $\emptyset$\\
\For{ $j$ in range($N-1$)}{ 
\For{ $0\leq i<j$}{ 
	Sample $J_{ij}$    \\
	Add $J_{ij}$ to Interactions \\ }
}
Initialize $\bm{\beta}$, $\bm{\gamma}$ and $\bm{\Omega}$ as arrays of length $p$\\

  \SetKwProg{myproc}{Function}{}{}

\myproc{LocalExpectationValue($\rho$,$O$)}{
	RCC $\leftarrow$ Select all gates in $\rho$ within the light cone of $O$ \\
	Find efficient contraction path to compute $\Tr(\rho O)$ restricted to RCC\\
	Perform contraction according to the contraction path found \\
	\Return $\Tr(\rho O)$ } 

\myproc{MinimizeEnergy(Interactions, $\bm{\beta}$, $\bm{\gamma}$, $\bm{\Omega}$)}{
\myproc{CalculateEnergy($\bm{\beta}$, $\bm{\gamma}$, $\bm{\Omega}$)}{
$\rho$ $\leftarrow$  Build the parity encoded QAOA circuit $ \ket{\psi(\bm{\gamma},\bm{\beta}, \bm{\Omega})} \bra{\psi(\bm{\gamma},\bm{\beta}, \bm{\Omega})}$ according to Eq.~\eqref{eq:psi} \\
energy = 0 \\
  \For{$O$ in Hamiltonian}{ 
	energy += LocalExpectationValue($\rho$,$O$) \\
}
\Return energy }

	energy, $\bm{\gamma}^{\star},\bm{\beta}^{\star}, \bm{\Omega}^{\star}$  $\leftarrow$ Minimize the energy of the Hamiltonians~\eqref{eq:H_PC} or~\eqref{eq:H_ST} \\
	\Return energy,  $\bm{\gamma}^{\star},\bm{\beta}^{\star}, \bm{\Omega}^{\star}$  }

energy, $\bm{\gamma}^{\star},\bm{\beta}^{\star}, \bm{\Omega}^{\star}$   $\leftarrow$ \textit{MinimizeEnergy(Interactions, $\bm{\beta}$, $\bm{\gamma}$, $\bm{\Omega}$)}

   \KwOutput{energy, $\bm{\gamma}^{\star},\bm{\beta}^{\star}, \bm{\Omega}^{\star}$   }
\end{small}
\end{algorithm}

To make a connection with the literature on simulating QAOA circuits with tensor networks~\cite{Streif2020}, we point out some differences with respect to paradigmatic Maxcut QAOA on $D$-regular graphs $G(N,D)$. However, once again we should note that the fixed $p$ algorithms are not directly comparable, for this we should first fix a performance ratio and then adjust $p$ in the parity-encoded algorithm such that fixed performance is reached.

It is a well known fact that in the large $N$ limit for Maxcut QAOA, the bulk of the RCC subgraphs are trees, followed by a first order contribution of single cycle subgraphs~\cite{Farhi2014}. Tree structures can be efficiently classically simulated. Indeed, it can be demonstrated that the contraction cost is at most scaling as $\mathcal{O}(\exp(p\cdot \texttt{tw}(G')))$, where $\texttt{tw}(G')$ denotes the treewith of the subgragh $G' \subset G$. For tree subgraphs, by default $\texttt{tw}(G')=1$, hence the classical scaling is only exponential in $p$, which allowed for efficient classical evaluations of the Maxcut QAOA cost function for up to $p=11$ for $D=3$, see e.g. Ref.~\cite{Streif2020}. (A tree RCC corresponding to this depth contains maximally $N'=8190$ qubits, as given by $2 \cdot \frac{(D-1)^{p+1}-1}{D-1}$. We plot these numbers in Fig.~\ref{fig:RCC_counting} as a function of $p$ for $D=3,4$.)

On the other hand, for parity-encoded QAOA, the RCCs all have the same square lattice topology (see Fig.~\ref{fig:RCCs} up to boundary effects), due to the locality of the algorithm. This means that the number of qubits will only grow quadratically with $p$, as maximally $(2p+2)^2$ (see Fig.~\ref{fig:RCC_counting}). However this reduced scaling in the number of qubits inside the RCC, does not lead to any simplification in the calculation. On the contrary, as the square lattice structure of the RCCs is now inherently loopy. (The treewidth of a square lattice graph is just the lattice width.) The cost of doing exact calculations now scales exponentially in this qubit count. The size of the largest tensor in the contraction path is then also maximally $2^{(2p+2)^2}$. Exponents up to $\sim 30$ can be simulated with a standard laptop. This strongly limits the exact calculation capabilities, either to very low $p$, or to a very small system sizes $N$ that does not span complete RCCs. For instance, the parity encoding of the $N=6$ graph shown in Fig.~\ref{fig:lhz_mapping} does not span a full reverse causal cone, even not at lowest depth $p=1$ (see Fig.~\ref{fig:RCCs}).

\newpage
\bibliographystyle{unsrtnat}
\bibliography{biblio}

\end{document}